\documentclass[twocolumn,english,showpacs,aps,prb,superscriptaddress,eqsecnum]{revtex4}
\usepackage[T1]{fontenc}
\usepackage[latin1]{inputenc}
\usepackage{amsmath}
\usepackage{graphicx}
\usepackage{amssymb}

\makeatletter
\usepackage{pslatex}
\usepackage{amssymb}
\usepackage{times}
\usepackage{amsmath}
\usepackage{graphics}
\usepackage{hyperref}
\hypersetup{backref, 
colorlinks=true,
citecolor=blue,
linkcolor=blue,
filecolor=blue,
urlcolor=blue
}

\usepackage{babel}
\makeatother
\begin{document}

\title{Correlation amplitude and entanglement entropy in random spin chains}

\author{Jos\'{e} A. Hoyos}

\email{hoyosj@umr.edu}

\affiliation{Instituto de F\'{\i}sica Gleb Wataghin, Unicamp, Caixa Postal 6165,
13083-970. Campinas, São Paulo, Brazil}

\affiliation{Department of Physics, University of Missouri-Rolla, Rolla, Missouri
65409, USA}

\author{Andr\'{e} P. Vieira}

\email{apvieira@ufc.br}

\affiliation{Departamento de Engenharia Metal\'{u}rgica e de Materiais, Universidade
Federal do Cear\'{a}, 60455-760, Fortaleza, Ceará, Brazil}

\author{N. Laflorencie}

\email{nicolas.laflorencie@epfl.ch}

\affiliation{Institute of Theoretical Physics, \'{E}cole Polytechnique F\'{e}d\'{e}rale
de Lausanne, CH-1015 Lausanne, Switzerland}

\affiliation{Department of Physics and Astronomy, University of British Columbia,
Vancouver, British Columbia, Canada, V6T 1Z1}

\author{E. Miranda}

\email{emiranda@ifi.unicamp.br}

\affiliation{Instituto de F\'{\i}sica Gleb Wataghin, Unicamp, Caixa Postal 6165,
13083-970. Campinas, São Paulo, Brazil}

\begin{abstract}
Using strong-disorder renormalization group, numerical exact diagonalization,
and quantum Monte Carlo methods, we revisit the random antiferromagnetic
XXZ spin-1/2 chain focusing on the long-length and ground-state behavior
of the average time-independent spin-spin correlation function $C(l)=\upsilon l^{-\eta}$.
In addition to the well-known universal (disorder-independent) power-law
exponent $\eta=2$, we find interesting universal features displayed
by the prefactor $\upsilon=\upsilon_{{\rm o}}/3$, if $l$ is odd,
and $\upsilon=\upsilon_{{\rm e}}/3$, otherwise. Although $\upsilon_{{\rm o}}$
and $\upsilon_{{\rm e}}$ are nonuniversal (disorder dependent) and
distinct in magnitude, the combination $\upsilon_{{\rm o}}+\upsilon_{{\rm e}}=-1/4$
is universal if $C$ is computed along the symmetric (longitudinal)
axis. The origin of the nonuniversalities of the prefactors is discussed
in the renormalization-group framework where a solvable toy model
is considered. Moreover, we relate the average correlation function
with the average entanglement entropy, whose amplitude has been recently
shown to be universal. The nonuniversalities of the prefactors are
shown to contribute only to surface terms of the entropy. Finally,
we discuss the experimental relevance of our results by computing
the structure factor whose scaling properties, interestingly, depend
on the correlation prefactors. 
\end{abstract}


\maketitle

\section{Introduction}

Random low-dimensional quantum spin systems have been intensively
investigated recently. The interplay between disorder, quantum fluctuations,
and correlations generates low-temperature phase diagrams with exotic
phases.~\cite{igloi-review} In this context, one of the most investigated
systems is the random antiferromagnetic (AF) quantum XXZ spin-$1/2$
chain, whose Hamiltonian reads \begin{equation}
H=\sum_{i}J_{i}\left(S_{i}^{x}S_{i+1}^{x}+S_{i}^{y}S_{i+1}^{y}+\Delta_{i}S_{i}^{z}S_{i+1}^{z}\right)~,\label{eq:H-heisenberg}\end{equation}
 in which $i$ labels the chain sites, $\mathbf{S}_{i}$ are the usual
spin-$1/2$ operators, $J_{i}$'s are positive uncorrelated random
variables drawn from a probability distribution $P_{0}\left(J\right)$,
and $\Delta_{i}$'s are anisotropy parameters, also random uncorrelated
variables.

The clean system, $J_{i}\equiv1$ and $\Delta_{i}\equiv\Delta$, is
a Tomonaga-Luttinger liquid for $-1<\Delta\leq1$, with well-known
asymptotic ground-state correlation functions,~\cite{luther-peschel-xxz,haldane-conjecture}\begin{align}
C_{{\rm c}}^{xx}\left(l\right) & =\left\langle S_{i}^{x}S_{i+l}^{x}\right\rangle =\left(-1\right)^{l}Fl^{-\eta_{{\rm c}}}-\tilde{F}l^{-\eta_{{\rm c}}-1/\eta_{{\rm c}}}~,\label{eq:cx-clean}\\
C_{{\rm c}}^{zz}\left(l\right) & =\left\langle S_{i}^{z}S_{i+l}^{z}\right\rangle =\left(-1\right)^{l}Al^{-1/\eta_{{\rm c}}}-\frac{1}{4\pi^{2}\eta_{{\rm c}}l^{2}}~,\label{eq:cz-clean}\end{align}
 as $l\rightarrow\infty$. The clean-system exponent is~\cite{luther-peschel-xxz}
$\eta_{{\rm c}}=1-\left(\arccos\,\Delta\right)/\pi$. At the {}``free-fermion''
point $\Delta=0$, the prefactors of the leading terms are known exactly,~\cite{lieb-schultz-mattis,mccoy-68}
being given by $A=1/\left(2\pi^{2}\right)$ and $F\approx0.147\,09$.~\cite{foot-mccoy}
Away from this point ($|\Delta|<1$), analytical forms for $A$ and
$F$ were derived by Lukyanov and Zamolodchikov~\cite{lukyanov-zamolodchikov-97,lukyanov-prefactor}
and checked numerically later on.~\cite{hikihara-furusaki-98} Furthermore,
the constant $\tilde{F}$, evaluated numerically in Ref.~\onlinecite{hikihara-furusaki-98},
is at least 1 order of magnitude smaller than $F$. At the isotropic
point $\Delta=1$, irrelevant operators become marginal, yielding
logarithmic corrections~\cite{affleck-prefactor,lukyanov-npb-98}\begin{equation}
C_{{\rm c}}^{xx}\left(l\right)=C_{{\rm c}}^{zz}\left(l\right)=\left(-1\right)^{l}\frac{\sqrt{\ln\, l}}{\left(2\pi\right)^{3/2}l}~.\label{eq:c-clean-xxx}\end{equation}

For $\Delta>1$, a spin gap opens and the system enters an antiferromagnetic
Ising phase; otherwise, for $\Delta<-1$, the chain becomes a gapped
Ising ferromagnet.

Disorder strongly modifies the behavior in the clean critical regime.
It was shown that even the least amount of disorder in $J_{i}$ destabilizes
the Tomonaga-Luttinger phase, provided $-1/2<\Delta_{i}\leq1$.~\cite{doty-fisher}
For $-1<\Delta_{i}\leq-1/2$, a finite amount of disorder is required
to destabilize the clean phase. The low-energy behavior of the random
AF spin-$1/2$ chain then corresponds to a random-singlet phase, characterized
by activated dynamical scaling with a \emph{universal} {}``tunneling''
exponent $\psi=1/2$, i.e., length ($l$) and energy ($\Omega$) scales
are related through $\Omega\sim\exp\left(-l^{\psi}\right)$, irrespective
of $P_{0}\left(J\right)$.~\cite{fisher94-xxz} Moreover, the transverse
and the longitudinal \emph{mean} spin-spin correlation functions decay
as a power law $\sim\upsilon l^{-\eta}$ for large distances, both
with the same universal exponent $\eta=2$.~\cite{doty-fisher,fisher94-xxz}
The mean value of the correlation function is dominated by rare widely
separated spin pairs coupled in strongly-correlated singlet states.
The remarkable fact that all correlations ($xx$, $yy$ and $zz$)
decay with the same exponent, irrespective of $\Delta$, can be ascribed
to the isotropy of the singlet state. In contrast, the \emph{typical}
value of the correlation function decays as a stretched exponential
$\sim\exp\left(-l^{\psi}\right)$. These results were obtained by
using the most successful theoretical tool to investigate such systems,
the real-space strong-disorder renormalization-group (SDRG) method,
first introduced in Refs.~\onlinecite{MDH-PRL} and \onlinecite{MDH-PRB}.

The main idea behind the SDRG method is to gradually lower the energy
scale by successively coupling the most strongly interacting spin
pairs into singlet states. At each step of the renormalization transformation,
one such pair is decimated out of the chain, and its remaining neighboring
spins become connected by a weaker renormalized coupling constant,
calculated within perturbation theory. Thus, in this framework, the
ground state can be viewed as a collection of {}``noninteracting''
singlets formed by arbitrarily distant spin pairs. Although this description
is not strictly exact, spin pairs do couple in states arbitrarily
close to singlets.~\cite{hoyos-rigolin}

Recently, efforts to compute certain numerical prefactors on disordered
systems have been made. Fisher and Young~\cite{fisher-young-RTFIM}
have shown that the end-to-end correlation amplitude of the random
transverse-field Ising chain at criticality is nonuniversal because
of some high-energy small-scale features that are not treated correctly
by the SDRG method. It is reasonable to expect that the same holds
for bulk correlations. Indeed, no sign of universality was found in
the correlation amplitude of the random XXZ chain.~\cite{henelius-girvin,laflorencie-correlacao-PRL,laflorencie-correlacao-PRB}

Refael and Moore,~\cite{refael-moore} on the other hand, have considered
the mean entanglement entropy $S\left(l\right)$ (for a recent review,
see Ref.~\onlinecite{review-amico-07}) between two complementary
subsystems $A$ (of size $l$) and $B$. Similarly to the clean system,~\cite{holzhey-larsen-wilczek,vidal-etal-prl03,calabrese-cardy-jstatmech04}
they have shown that $S\left(l\right)=b+\left(\gamma/3\right)\ln\, l$,
diverging logarithmically with the subsystem size. More interestingly,
the prefactor $\gamma$ is universal for a large class of systems
governed by an infinite-randomness fixed point, namely, the random
transverse-field Ising chain at the critical point and the spin-1/2
random antiferromagnetic XXZ chain. Later, this amplitude was shown
to be universal for a broader class of systems governed by infinite-randomness
fixed points: the random $q$-state Potts chain and the $Z_{q}$ clock
chain~\cite{santachiara-06} and the random antiferromagnetic spin-$S$
chain at the random-singlet phase.~\cite{saguia-entanglement-07,refael-moore-07}
Moreover, it has also been shown that this amplitude is also universal
in a large class of aperiodic chains.~\cite{juhasz-zimboras,igloi-juhasz-zimboras-epl07}
In the renormalization-group sense, and following Fisher and Young,~\cite{fisher-young-RTFIM}
Refael and Moore~\cite{refael-moore} argued that the nonuniversalities
of the correlation amplitudes are related to inaccuracies of order
of the lattice spacing in the location of the effective spins. Such
errors can only contribute a surface term $b$ to the entanglement
entropy, and therefore, its prefactor $\gamma$ should remain universal.
Notably, this should explain why all those other models displaying
a random-singlet-like ground state show universal entropy prefactors.

Our aim in this work is to further explore the issue of universality
in the behavior of ground-state correlation functions in random antiferromagnetic
XXZ chains and make some direct links between spin correlations, structure
factor, and entanglement entropy. We first calculate exactly, \emph{within}
the SDRG framework, the numerical prefactor $\upsilon$ of the mean
correlation function $\overline{\left\langle \mathbf{S}_{i}\cdot\mathbf{S}_{i+l}\right\rangle }$
in the limit $l\rightarrow\infty$, by relating it to the distribution
of singlet-pair bond lengths in the ground state. Surprisingly, it
turns out to be \emph{universal} and equals $\upsilon_{{\rm o}}=-1/4$,
if $l$ is odd, and $\upsilon_{{\rm e}}=0$, otherwise, because the
{}``noninteracting'' singlets can only be formed between spins separated
by an odd number of lattice sites. Naturally, this result is an artifact
of the (perturbative) SDRG scheme, as shown by exact diagonalization
(ED) studies of the XX model~\cite{henelius-girvin,laflorencie-correlacao-PRL}
(in which $\Delta_{i}=0$, $\forall i$) and quantum Monte Carlo (QMC)
calculations applied to the isotropic Heisenberg model~\cite{laflorencie-correlacao-PRB}
(in which $\Delta_{i}=1$, $\forall i$). Nevertheless, as we show
from ED calculations, in the XX limit, the long-distance behavior
of the longitudinal mean correlation function $\overline{\left\langle S_{i}^{z}S_{i+l}^{z}\right\rangle }$
is shown to be very well described by this renormalization-group prediction,
while the transverse mean correlation function $\overline{\left\langle S_{i}^{x}S_{i+l}^{x}\right\rangle }$
exhibits two distinct prefactors, $\upsilon_{{\rm o}}^{x}$ and $\upsilon_{{\rm e}}^{x}$
for odd and even $l$, respectively {[}see Eq.~(\ref{eq:C-transversal}){]}.

Furthermore, we explicitly show that the mean entanglement entropy
is directly related to the bond-length distribution of the singlet
pairs and, therefore, directly related to the correlation function.
This is interesting because it links a pairwise quantity (correlation)
with a blockwise one (entropy).

Since such a relation arises in the scenario of {}``noninteracting''
spin singlets, in which nonuniversal effects are omitted, we introduce
a toy model in which the correlations between different singlet pairs
can be treated \emph{exactly}. From the toy model, we gain some insight
into the microscopic nature of the random-singlet phase and quantify
the role of the interactions between the singlets.

The picture emerging from our analytical results, and confirmed by
our own ED and QMC calculations, is the following. At long-length
scales, the chain can be recast as a collection of noninteracting
\emph{effective-spin} singlets sharing strong pairwise correlations.
These effective spins are clusters of original spin variables whose
number depends on the details of the coupling constant distribution.
With respect to the original spin variables, the singletlike correlations
{}``spread'' among the spins in the cluster, an effect which leads
to the nonuniversal high-energy contributions discussed in the literature.~\cite{fisher-young-RTFIM,refael-moore}
Here, we go further by quantifying these contributions within the
exactly solvable toy model. Interestingly, whenever the correlation
is computed along a symmetry ($z$) axis, it equals the corresponding
-1/4 singlet contribution, i.e., summing the correlations between
all pairs of spins sitting at different clusters gives -1/4. For the
mean correlation function, the result is that the combination of prefactors
$\upsilon_{{\rm o}}^{z}+\upsilon_{{\rm e}}^{z}=-1/4$ is \emph{universal}.
For correlations along a nonsymmetry ($x$ or $y$) axis, not only
the prefactors are nonuniversal but also the combination $\upsilon_{{\rm o}}^{x}+\upsilon_{{\rm e}}^{x}$.
This points to the importance of symmetry for the observed universality,
a feature which was absent from the previously considered models.~\cite{fisher-young-RTFIM}
Finally, effective spins contribute only nonuniversal surface terms
to the entanglement entropy, as expected. When one traces completely
one of the spin clusters, its contribution to the entanglement entropy
is the same as that of a singlet pair; only when the boundary between
the subsystems is crossed by one of the clusters does that cluster
contribute a nonuniversal term. Therefore, the entanglement-entropy
prefactor is universal regardless of the existence of a symmetry axis.

As a supplement, we compute both analytically and numerically the
static structure factor ${\cal S}\!\left(q\right)$, which can be
probed by neutron scattering experiments. Interestingly, we show that
${\cal S}\!\left(q\right)$ is dictated by disorder in the small-$q$
limit, namely, ${\cal S}\!\left(q\ll1\right)=\kappa\left|q\right|$.
This is a consequence of two facts: (i) the decay exponent $\eta=2$
being universal, and (ii) the magnitude of the numerical prefactors
$\upsilon_{{\rm o}}$ and $\upsilon_{{\rm e}}$ being different. Moreover,
we find that $\kappa=-\pi^{2}(\upsilon_{{\rm o}}+\upsilon_{{\rm e}})/3$,
which implies that $\kappa=\pi^{2}/12$ is universal for ${\cal S}$
computed along a symmetry axis. On the other hand, the behavior near
the AF peak $q=\pi$ is dominated by the characteristic divergence
of the clean system. However, the true divergence at $q=\pi$ is suppressed
by disorder and the peak width is broadened. Since there is no divergence
in the case of ${\cal S}\!\left(q\right)$ along the $z$ axis in
the XX model, disorder universally determines its behavior near the
AF peak, i.e., ${\cal S}^{z}\left(q=\pi-\epsilon\right)=\pi-\kappa\left|\epsilon\right|$
for $\epsilon\ll1$. Only for $q\approx\pi/2$ is the clean-system
behavior ${\cal S}^{z}\left(q\right)=\left|q\right|$ found.

The remainder of this paper is as follows. We derive the universal
SDRG expression for the mean correlation function in Sec.~\ref{sec:Mean-corr-func},
reporting our numerical analyses in Sec.~\ref{sec:Numerical-results}.
Section \ref{sec:An-exactly-solvable} discusses an exactly solvable
model that yields instructive results on the origin of the universal
behavior of correlation functions. In Sec.~\ref{sec:Entanglement-entropy},
we derive the entanglement entropy and relate it to the distribution
of singlet lengths and to the correlation function. We discuss the
experimental relevance of our results by computing the structure factor
in Sec.~\ref{sec:The-structure-factor}. Finally, we make some concluding
remarks in Sec.~\ref{sec:Conclusions}.

\section{Mean correlation function in the strong-disorder renormalization-group
framework\label{sec:Mean-corr-func}}

We start this section with a brief review of the SDRG method, followed
by the derivation of the mean correlation function.

\subsection{Strong-disorder renormalization-group method: A brief review}

\begin{figure}
\begin{center}\includegraphics[%
  clip,
  width=0.7\columnwidth,
  keepaspectratio]{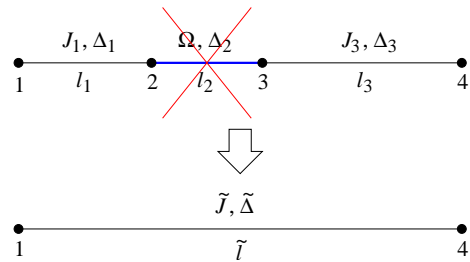}\end{center}

\caption{\label{cap:decimation-procedure}(Color online) Schematic decimation
procedure (see text).}
\end{figure}

The main idea behind the SDRG method is to reduce the energy scale
by integrating out the strongest couplings and renormalizing the remaining
ones. In the present case, one locates the strongest coupling constant
$\Omega=\max\left\{ J_{i}\right\} $, say, $J_{2}$, and then exactly
treats the two-spin Hamiltonian $H_{0}=\Omega\left(S_{2}^{x}S_{3}^{x}+S_{2}^{y}S_{3}^{y}+\Delta_{2}S_{2}^{z}S_{3}^{z}\right)$,
considering $H_{1}=H-H_{0}$ as a perturbation.~\cite{MDH-PRL} At
low energies, spins $S_{2}$ and $S_{3}$ {}``freeze'' into a (nonmagnetic)
singlet state, with the result that they can be effectively removed
from the chain, provided that the neighboring spins $S_{1}$ and $S_{4}$
are now connected by a renormalized coupling constant\begin{equation}
\tilde{J}=\frac{J_{1}J_{3}}{\left(1+\Delta_{2}\right)\Omega}~,\label{eq:j-tilde}\end{equation}
 calculated within second-order perturbation theory. The anisotropy
parameter is also renormalized to $\tilde{\Delta}=\Delta_{1}\Delta_{3}(1+\Delta_{2})/2$.
Note that $\tilde{J}$ is smaller than either $J_{1}$, $J_{3}$,
or $\Omega$, leading to an overall decrease in the energy scale.
After the decimation procedure, the distance between $S_{1}$ and
$S_{4}$, which are now nearest neighbors, is renormalized to \begin{equation}
\tilde{l}=l_{1}+l_{2}+l_{3}~,\label{eq:l-tilde}\end{equation}
 with $l_{i}$ defined as the distance between the spin $S_{i}$ and
its nearest neighbor to the right. The SDRG decimation scheme is illustrated
in Fig.~\ref{cap:decimation-procedure}.

Clearly, as the SDRG scheme is iterated and the energy scale $\Omega$
is reduced, the distribution of effective coupling constants $P_{J}(J;\Omega)$
is modified. Searching for fixed-point coupling-constant distributions
$P_{J}^{*}\left(J;\Omega\right)$, Fisher found that there is only
one regular stable fixed point\begin{equation}
P_{J}^{*}\left(J;\Omega\right)=\theta\left(J\right)\theta\left(\Omega-J\right)\frac{\alpha}{\Omega}\left(\frac{\Omega}{J}\right)^{1-\alpha},\label{eq:P*J-Fisher}\end{equation}
 in which $\theta\left(x\right)$ is the Heaviside step function and
with~\cite{fisher94-xxz} $\alpha=-1/\ln\,\Omega$ (we set the initial
energy scale $\Omega_{0}$ to 1). This has been named an infinite-randomness
fixed point (IRFP) since, as $\Omega\rightarrow0$, the distribution
becomes infinitely broad, i.e., $\sqrt{\textrm{Var }J}/\overline{J}\rightarrow\infty$,
where Var $J$ and $\bar{J}$ are the variance and mean value of the
coupling constants, respectively. Thus, the perturbative decimation
procedure becomes more and more precise along the flow because the
probability that both $J_{1}$ and $J_{3}$ are much smaller than
$J_{2}$ increases as the energy scale is lowered.

A useful quantity to be calculated is the fraction $n_{\Omega}$ of
{}``active'' (not yet decimated) spins at the energy scale $\Omega$.
It is obtained from the rate equation \begin{equation}
{\rm d}n_{\Omega}=2n_{\Omega}P_{J}\left(J=\Omega;\Omega\right){\rm d}\Omega~,\label{eq:rate}\end{equation}
 where $2n_{\Omega}P_{J}\left(J=\Omega;\Omega\right){\rm d}\Omega$
is the fraction of decimated spins when the energy scale is lowered
from $\Omega$ to $\Omega-{\rm d}\Omega$. Hence, close to the fixed
point, \begin{equation}
n_{\Omega}\sim\frac{1}{\ln^{2}\Omega}~.\label{eq:n-omega-approx}\end{equation}

Equation~(\ref{eq:n-omega-approx}) directly gives the low-temperature
magnetic susceptibility $\chi\left(T\right)$. One iterates the SDRG
procedure until the low-temperature scale $T$. Spin pairs decimated
at high energy scales $\Omega\gg T$ are {}``frozen'' into singlet
states, and thus their contribution to the magnetic susceptibility
can be neglected. As the fixed-point distribution is very broad, all
couplings between active spins are very weak compared to $T$, and
the active spins can be considered as essentially free spins, each
contributing a Curie term to the susceptibility.~\cite{fisher94-xxz}
Therefore, \begin{equation}
\chi\sim\frac{n_{T}}{T}\sim\frac{1}{T\ln^{2}T}~.\label{eq:mag-suscep-approx}\end{equation}

The low-energy modes are also given by Eq.~(\ref{eq:n-omega-approx}).
These modes are polarizations of widely separated weakly coupled singlet
pairs, decimated at the energy scale $\Omega$ for which the mean
distance between spins was $l\sim n_{\Omega}^{-1}$. Thus, the energy
cost $\Omega$ to break a singlet of length $l$ is\begin{equation}
\Omega\sim\exp\left(-l^{\psi}\right)~,\label{eq:omega-L}\end{equation}
 in which $\psi=1/2$.~\cite{fisher94-xxz} This unusual exponential
relation between $\Omega$ and $l$ is named {}``activated'' dynamical
scaling and $\psi$ has been dubbed the {}``tunneling exponent.''

The scaling behavior of the mean correlation function $C\left(l\right)$
is cleverly obtained when one realizes that \emph{typically} two very
distant spins are not in a singlet state and thus are only weakly
correlated. On the other hand, some rare and arbitrarily separated
spin pairs that were decimated together are strongly correlated and
hence dominate the long-distance behavior of $C\left(l\right)$. Therefore,
the mean correlation function must be proportional to the total number
of spin singlets decimated at the length scale $l$. Since the probability
of decimating a spin pair is proportional to the probability that
both spins have not been decimated yet, it follows that\begin{equation}
C\left(l\right)\sim\left(-1\right)^{l}n_{\Omega}^{2}\sim\frac{\left(-1\right)^{l}}{l^{\eta}}~,\label{eq:C-Fisher}\end{equation}
 with $\eta=2$.~\cite{fisher94-xxz}

In contrast, the typical correlation function $C_{{\rm typ}}\left(l\right)$
behaves quite differently. Its long-distance behavior is obtained
by the following argument. Suppose spins $S_{2}$ and $S_{3}$ are
those to be decimated at a given SDRG step, as in Fig.~\ref{cap:decimation-procedure}.
In that case, the correlation between $S_{2}$ and $S_{3}$ equals
$-3/4+\mathcal{O}\!\left(J_{3}/\Omega\right)^{2}$, while the correlation
between $S_{4}$ and $S_{3}$ is of order $-J_{3}/\Omega$. Thus,
the typical value of the correlation function will be proportional
to the typical value of $\tilde{J}/\Omega$. Using the fixed-point
distribution (\ref{eq:P*J-Fisher}), one finds $\ln\left|C_{{\rm typ}}\left(l\right)\right|\sim-\iota\sqrt{l}$,
i.e., the typical correlation function decays as a stretched exponential,
with a nonuniversal prefactor $\iota$ of order unity.~\cite{fisher94-xxz}

\subsection{Mean correlation function\label{sub:The-mean-correlation}}

We now derive the mean correlation function in a more formal calculation
which allows us to compute its amplitude in addition to its power-law
decay. In the SDRG framework, the mean correlation function $C(l)$
between spins separated by a distance $l$ is obtained from the corresponding
distribution of singlet-pair bond lengths in the ground state $P_{{\rm s}}(l)$,\begin{equation}
C(l)=-\frac{3}{8}P_{{\rm s}}(l)~,\end{equation}
 since each singlet contributes a factor of $-3/4$ to $C(l)$ and
there are two spins in each singlet.

The singlet-bond length distribution $P_{{\rm s}}(l)$ can be calculated
from\begin{equation}
P_{{\rm s}}\left(l\right)=2\int_{0}^{\Omega_{0}}n_{\Omega}P\left(J=\Omega,l;\Omega\right){\rm d}\Omega~,\label{eq:Ps-integral}\end{equation}
 where $P\left(J,l;\Omega\right){\rm d}J{\rm d}l$ is the probability
of finding a coupling constant between $J$ and $J+{\rm d}J$ connecting
spins separated by a distance between $l$ and $l+{\rm d}l$ at the
energy scale $\Omega$, and the factor $2$ comes from normalization.
If we follow exactly the joint probability $P\left(J,l;\Omega\right)$
along the SDRG flow, then we can obtain an exact expression for $P_{{\rm s}}(l)$.
In fact, we only need $P\left(J,l;\Omega\right)$ at $J=\Omega$.

It turns out that we can carry out this task for $\Delta_{i}\equiv0$
and couplings taken from the family of initial distributions,\begin{equation}
P_{0}\left(J\right)=\theta\left(J\right)\theta\left(\Omega_{0}-J\right)\frac{\vartheta_{0}}{\Omega_{0}}\left(\frac{\Omega_{0}}{J}\right)^{1-\vartheta_{0}},\label{eq:dist-inc}\end{equation}
 in which $\vartheta_{0}>0$ gauges the strength of the initial disorder
and $\Omega_{0}$ sets the initial energy scale.~\cite{foot-igloi}
We first calculate the density of active spins $n_{\Omega}$. For
that, we need $P_{J}\left(J;\Omega\right)=\int P\left(J,l;\Omega\right){\rm d}l$,
which is obtained from the flow equation~\cite{MDH-PRL}\begin{align}
-\frac{\partial P_{J}}{\partial\Omega}= & P_{J}(\Omega;\Omega)\int{\rm d}J_{1}{\rm d}J_{3}P_{J}\left(J_{1};\Omega\right)P_{J}\left(J_{3};\Omega\right)\nonumber \\
 & \qquad\times\delta\left(J-\frac{J_{1}J_{3}}{\Omega}\right)~.\label{eq:PJ-flux}\end{align}

Introducing the \emph{Ansatz}~\cite{igloi-det-z-PRL,igloi-det-z-PRB}\begin{equation}
P_{J}\left(J;\Omega\right)=\frac{\vartheta\left(\Omega\right)}{\Omega}\left(\frac{\Omega}{J}\right)^{1-\vartheta\left(\Omega\right)}\label{eq:ansatz-PJ}\end{equation}
 into Eq.~(\ref{eq:PJ-flux}) yields\begin{equation}
\vartheta\left(\Omega\right)=\frac{\vartheta_{0}}{1+\vartheta_{0}\Gamma}~,\end{equation}
 where $\Gamma=\ln\left(\Omega_{0}/\Omega\right)$. Thus, from the
rate equation (\ref{eq:rate}), we obtain\begin{equation}
n_{\Omega}=\frac{1}{\left(1+\vartheta_{0}\Gamma\right)^{2}}~,\label{eq:n-omega-exact}\end{equation}
 and Eq.~(\ref{eq:n-omega-approx}) is recovered in the low-energy
limit $\Gamma\rightarrow\infty$.

We now need to follow the SDRG flow of the joint distribution $P\left(J,l;\Omega\right)$,
which is governed by the equation~\cite{fisher94-xxz}\begin{align}
\frac{\partial P}{\partial\Omega}= & -\int{\rm d}l_{1}{\rm d}l_{2}{\rm d}l_{3}{\rm d}J_{1}{\rm d}J_{3}P\left(J_{1},l_{1}\right)P\left(\Omega,l_{2}\right)P\left(J_{3},l_{3}\right)\nonumber \\
 & \times\delta\left(l-l_{1}-l_{2}-l_{3}\right)\delta\left(J-\frac{J_{1}J_{3}}{\Omega}\right)~.\label{eq:flow-joint-dist}\end{align}
 As shown in Appendix~\ref{sec:Calculation-of-P}, this can be done
exactly by Laplace transforming $P\left(J,l;\Omega\right)$ and using
an \emph{Ansatz} for the corresponding flow equation. The final result
for $P(\Omega,l)\equiv P\left(J=\Omega,l;\Omega\right)$ is\begin{equation}
P\left(\Omega,l\right)=\frac{4\pi^{2}}{\Omega a^{2}\Gamma^{3}}\sum_{n=1}^{\infty}\left(-1\right)^{n+1}n^{2}\exp\left\{ -\left(\frac{n\pi}{a\Gamma}\right)^{2}l\right\} ~,\label{eq:P-omega-l}\end{equation}
 where $a=\vartheta_{0}\sqrt{2l_{0}},$ and $l_{0}\equiv1$ is the
{}``bare'' lattice spacing. Although the leading term of Eq.~(\ref{eq:P-omega-l})
had been obtained before,~\cite{fisher94-xxz} the explicit dependence
on the initial disorder distribution encoded in $a$ was not emphasized.
As will be shown next, this dependence is essential for our discussion.

Plugging Eqs.~(\ref{eq:n-omega-exact}) and (\ref{eq:P-omega-l})
into Eq.~(\ref{eq:Ps-integral}), we obtain\begin{flalign}
P_{{\rm s}}\left(l\right) & =\frac{8\pi^{2}}{a^{2}}\sum_{n=1}^{\infty}\left(-1\right)^{n+1}n^{2}\int_{0}^{\infty}\frac{e^{-\left(\frac{n\pi}{a\Gamma}\right)^{2}l}}{\left(1+\vartheta_{0}\Gamma\right)^{2}\Gamma^{3}}{\rm d}\Gamma\nonumber \\
 & =8\frac{l_{0}}{l}\sum_{n=1}^{\infty}\frac{\left(-1\right)^{n+1}}{\pi^{2}n^{2}}f\left(l,n\right)\label{eq:Ps-before}\\
 & =\frac{2l_{0}}{3l^{2}}\left\{ 1+\mathcal{O}\left(\sqrt{l_{0}/l}\right)\right\} ~,\label{eq:Ps}\end{flalign}
 where \begin{equation}
f\left(l,n\right)=\frac{1}{l_{0}}\int_{0}^{\infty}\frac{e^{-\epsilon}{\rm d}\epsilon}{\left(\sqrt{2}/\left(\pi n\right)+\sqrt{l/\left(l_{0}\epsilon\right)}\right)^{2}}~,\end{equation}
 and we used $f\left(l\gg l_{0},n\right)\rightarrow1/l\left\{ 1+\mathcal{O}\left(\sqrt{l_{0}/l}\right)\right\} $
in the last step. As explicitly shown in Eq.~(\ref{eq:Ps-before}),
the distribution of singlets in the ground state is independent of
the initial disorder parameter $\vartheta_{0}$ at \emph{all} length
scales. Moreover, it follows a \emph{universal} power law in the large-distance
limit.

Finally, taking into account that singlets can only be formed between
spins separated by distances corresponding to odd multiples of $l_{0}$,
the mean correlation function takes the \emph{universal} form\begin{equation}
C_{{\rm u}}\left(l\right)=-\upsilon\left(\frac{l_{0}}{l}\right)^{2}\times\left\{ \begin{array}{cl}
1, & \textrm{if }l/l_{0}\textrm{ is odd},\\
0, & \textrm{otherwise},\end{array}\right.\label{eq:corr-universal}\end{equation}
 where $\upsilon=1/4$, irrespective of the initial disorder parameter.
Note that Eq.~(\ref{eq:corr-universal}) recovers Fisher's scaling
result Eq.~(\ref{eq:C-Fisher}). In view of the fact that correlations
between the spins in a singlet state are isotropic, correlations between
components of the spins along a given direction $\alpha=x$, $y$,
or $z$ should behave as \begin{equation}
C_{{\rm u}}^{\alpha\alpha}\left(l\right)=-\frac{1}{3}\upsilon\left(\frac{l_{0}}{l}\right)^{2}\times\left\{ \begin{array}{cl}
1, & \textrm{if }l/l_{0}\textrm{ is odd},\\
0, & \textrm{otherwise},\end{array}\right.\label{eq:corr-universal-aa}\end{equation}
 with a prefactor given by $-\upsilon/3=-1/12$.

\begin{figure}
\begin{center}\includegraphics[%
  clip,
  width=0.9\columnwidth,
  keepaspectratio]{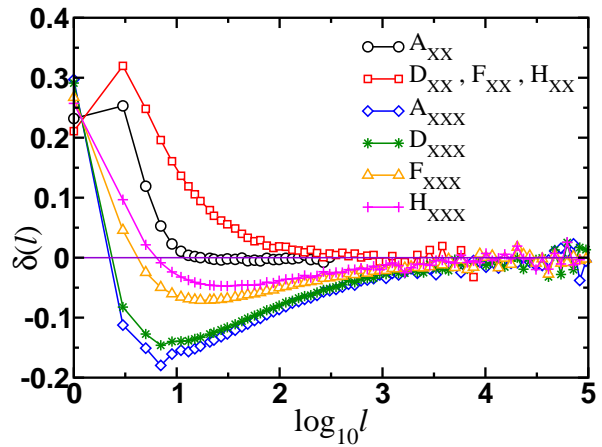}\end{center}

\caption{\label{cap:Relative-difference}(Color online) Relative difference
$\delta=C/C_{{\rm u}}-1$ between the mean correlation function $C$
and the universal prediction $C_{{\rm u}}$ as a function of the distance
$l$ in the SDRG framework, for various choices of initial disorder,
and both XX and isotropic Heisenberg (or XXX) models. Chains A, D,
F, and H have couplings distributed according to $P_{0}\left(J\right)$
{[}see Eq.~(\ref{eq:P0(Jmin,q0)}){]} with $\left(J_{{\rm min}},\vartheta_{0}\right)$
equal to $\left(0.5,1\right)$, $\left(0,3\right)$, $\left(0,1\right)$,
and $\left(0,0.3\right)$, respectively. The results for chains F$_{{\rm XX}}$
and H$_{{\rm XX}}$ (omitted for clarity) are statistically indistinguishable
from those for chain D$_{{\rm XX}}$. Error bars (not shown for clarity)
are of the order of the statistical data fluctuations. Lines are guides
to the eyes.}
\end{figure}

In order to check the prediction of Eq.~(\ref{eq:corr-universal}),
we calculated the mean correlation function $C\left(l\right)$ from
numerical implementations of the SDRG algorithm on very large chains
($2\times10^{7}$ sites), with initial couplings following probability
distributions of the form \begin{equation}
P_{0}\left(J\right)=\frac{\theta\left(J-J_{{\rm min}}\right)\theta\left(\Omega_{0}-J\right)}{1-\left(J_{{\rm min}}/\Omega_{0}\right)^{\vartheta_{0}}}\frac{\vartheta_{0}}{\Omega_{0}}\left(\frac{\Omega_{0}}{J}\right)^{1-\vartheta_{0}},\label{eq:P0(Jmin,q0)}\end{equation}
 where $\Omega_{0}=1$, $\vartheta_{0}>0$, and $J_{{\rm min}}\geq0$.
Figure~\ref{cap:Relative-difference} shows, for various chains,
the relative difference between the calculated correlation function
and the universal prediction, $\delta(l)=C\left(l\right)/C_{\textrm{u}}\left(l\right)-1$,
as a function of $l$. We considered both the XX ($\Delta_{i}\equiv0$)
and the isotropic Heisenberg ($\Delta_{i}\equiv1$) models, for which
we averaged over $100$ and $1000$ disorder realizations, respectively.
In agreement with the previous analysis, the long-distance behavior
of the correlation functions is well described by the universal prediction
$C_{\textrm{u}}\left(l\right)$, regardless of the model under consideration,
within an error of less than 5\%. Moreover, the mean correlation function
of chains D$_{{\rm XX}}$, F$_{{\rm XX}}$, and H$_{{\rm XX}}$ (all
of which have $J_{{\rm min}}=0$, as described in the figure caption)
are statistically identical at \emph{all} length scales, in agreement
with Eq.~(\ref{eq:Ps-before}), which predicts the same short-distance
behavior for those spin chains whose coupling constants are distributed
according to Eq.~(\ref{eq:dist-inc}). Notice that $\delta(l)$ approaches
zero for large $l$ even for distributions with $J_{\mathrm{min}}>0$,
which clearly do not belong to the particular class of distributions
{[}Eq.~(\ref{eq:dist-inc}){]} employed in the derivation of $C_{\mathrm{u}}(l)$.

The clear difference between the convergence rates of the mean correlation
functions in the XX and Heisenberg models is due to the extra numerical
prefactor of $1/2$ present in the recursion relation of the latter
model {[}cf. Eq.~(\ref{eq:j-tilde}){]}, which delays the convergence
of $C(l)$ to the asymptotic form $C_{{\rm u}}(l)$. This prefactor
(which becomes negligible as the SDRG scheme proceeds) alters the
relation between length and energy scales, relevant for the derivation
of $C_{\textrm{u}}$. However, at logarithmically large energy scales,
$\Gamma=\ln\left(\Omega_{0}/\Omega\right)\gg\ln\,2$, the simple relation
between length and energy scales in Eq.~(\ref{eq:omega-L}) is recovered.

\section{Numerical results\label{sec:Numerical-results}}

We now confront the predicted long-distance form of the mean correlation
function, given in Sec.~\ref{sec:Mean-corr-func}, with numerical
results for XX chains, obtained through the mapping to free fermions,
and for isotropic Heisenberg chains, obtained by quantum Monte Carlo
(QMC) calculations.

\subsection{XX chains\label{sub:XX-chains}}

\begin{figure}
\begin{center}\includegraphics[%
  clip,
  width=1\columnwidth,
  keepaspectratio]{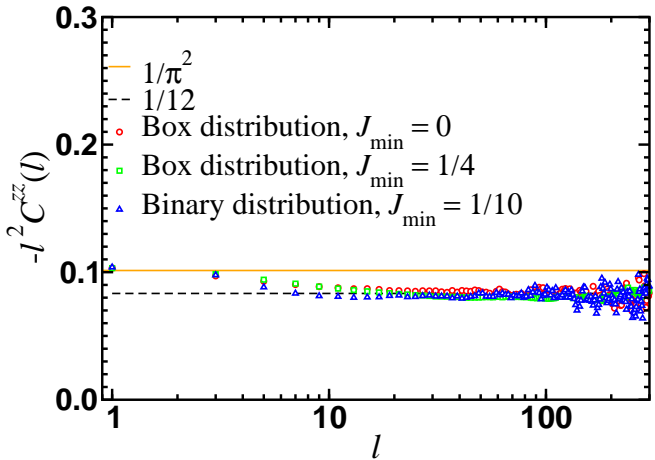}\end{center}

\caption{\label{cap:longitudinal-C}(Color online) Dependence of $-l^{2}C^{zz}\left(l\right)$
on the spin separation $l$ in the XX model for three different probability
distributions of the couplings: a box distribution ($\vartheta_{0}=1$)
with $J_{{\rm min}}=0$, a box distribution with $J_{{\rm min}}=1/4$,
and a binary distribution with $J_{{\rm min}}=1/10$ {[}see Eqs.~(\ref{eq:box})
and (\ref{eq:binary}){]}. The orange solid line corresponds to the
disorder-free prediction (Ref. \onlinecite{lieb-schultz-mattis})
$-l^{2}C_{{\rm c}}^{zz}=1/\pi^{2}$. For short-length scales, all
chains approach the behavior of the uniform system. After a disorder-dependent
crossover length, the data approach the universal prediction of Eq.~(\ref{eq:corr-universal-aa}),
which is indicated by the black dashed line. Statistical fluctuations
increase with $l$, and so results for $l\geq300$, as well as error
bars, are omitted for clarity.}
\end{figure}

We analyzed disordered XX chains with periodic boundary conditions,
and coupling constants following box distributions \begin{equation}
P_{0}\left(J\right)=\frac{\theta\left(J-J_{{\rm min}}\right)\theta\left(\Omega_{0}-J\right)}{1-\left(J_{{\rm min}}/\Omega_{0}\right)^{\vartheta_{0}}}\frac{\vartheta_{0}}{\Omega_{0}}\left(\frac{\Omega_{0}}{J}\right)^{1-\vartheta_{0}},\label{eq:box}\end{equation}
 with $\Omega_{0}=1$, $\vartheta_{0}>0$, and $J_{{\rm min}}>0$,
or binary distributions \begin{equation}
Q_{0}\left(J\right)=\frac{1}{2}\delta\left(J-J_{{\rm min}}\right)+\frac{1}{2}\delta\left(J-\Omega_{0}\right)~.\label{eq:binary}\end{equation}
 Below, we present results for different choices of parameters. Figure~\ref{cap:longitudinal-C}
shows the mean longitudinal correlation function $C^{zz}\left(l\right)=\overline{\left\langle S_{i}^{z}S_{i+l}^{z}\right\rangle }$
as a function of the spin separation $l$ for a chain with $4\,000$
sites and couplings taken from three probability distributions: two
boxlike distributions ($\vartheta_{0}=1$) with $J_{{\rm min}}=1/4$
and $J_{{\rm min}}=0$, and one binary distribution with $J_{{\rm min}}=1/10$,
in which we average over $700$, $1000$, and $800$ disorder realizations,
respectively. Other disorder distributions give similar results. The
short-length behavior approaches the uniform-system result,~\cite{lieb-schultz-mattis}
$C^{zz}\left(l\right)=-\left(\pi l\right)^{-2}$. After a disorder-dependent
crossover length,~\cite{laflorencie-correlacao-PRL,laflorencie-correlacao-PRB}
the mean longitudinal correlation function decays as a power law with
exponent $\eta=2$, and the prefactor clearly approaches the universal
value $-1/12$ {[}see Eq.~(\ref{eq:corr-universal-aa}){]}. Although
not shown in the figure, the typical longitudinal correlation function
$C_{\mathrm{typ}}^{zz}\left(l\right)$, in contrast, has a nonuniversal
prefactor.

It is remarkable how the random-singlet hallmark appears in Fig.~\ref{cap:longitudinal-C}.
Since $C^{zz}\sim l^{-2}$ both in the clean and in the disordered
cases, one could naively think that disorder does not play any role
for $C^{zz}$. However, our statistics are good enough to distinguish
the different prefactors. We stress that the fluctuations seen at
larger length scales reflect only the increasingly and inevitably
poorer statistics, since the number of singlet pairs decreases as
$l^{-2}$ and their relative fraction becomes smaller and smaller.
Indeed, the mean correlation function is self-averaging. This could
be directly double checked from the decrease of the relative fluctuations
with the square root of the inverse chain size.

\begin{figure}
\begin{center}\includegraphics[%
  clip,
  width=1\columnwidth,
  keepaspectratio]{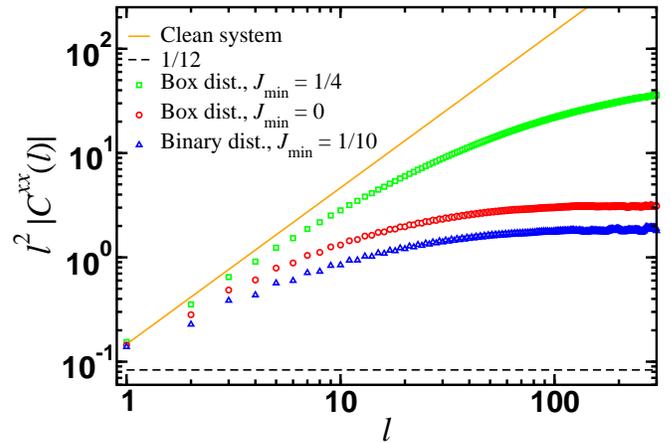}\end{center}

\caption{\label{cap:transverse-C}(Color online) Dependence of $l^{2}\left|C^{xx}\left(l\right)\right|$
on the spin separation $l$ for the XX model with the same coupling
distributions as in Fig.~\ref{cap:longitudinal-C}. The short-distance
behavior, as in the uniform system, corresponds to a power law with
exponent $\eta_{{\rm c}}=1/2$. The random-singlet nature of the ground
state governs the long-distance behavior, with $\eta=2$, but nonuniversal
prefactors. The orange solid line corresponds to the clean-system
transverse correlation function (Ref. \onlinecite{mccoy-68}) $\left(-1\right)^{l}0.147\,09/\sqrt{l}$.
The black dashed line is the prediction for the disordered system
given in Eq.~(\ref{eq:corr-universal-aa}) (see text). }
\end{figure}

We now turn our attention to the transverse mean correlation function
$C^{xx}\left(l\right)$. Compared to its longitudinal counterpart,
it behaves quite differently. The short-distance behavior, as in the
uniform system, corresponds to a power-law decay with exponent $\eta_{{\rm c}}=1/2$.
After a nonuniversal crossover length, the random-singlet behavior
is recovered, with a universal exponent $\eta=2$, but nonuniversal
prefactors, as shown in Fig.~\ref{cap:transverse-C} and previously
pointed out in Refs.~\onlinecite{henelius-girvin,laflorencie-correlacao-PRL,laflorencie-correlacao-PRB}.
Without loss of generality, the large-distance scaling form of $C^{xx}(l)$
can be written as \begin{equation}
C^{xx}\left(l\gg1\right)=\frac{1}{3}\left\{ \begin{array}{ll}
\upsilon_{{\rm o}}^{x}l^{-2}, & \textrm{if }l\textrm{ is odd,}\\
\upsilon_{{\rm e}}^{x}l^{-2}, & \textrm{otherwise,}\end{array}\right.\label{eq:C-transversal}\end{equation}
 with suitably chosen functions $\upsilon_{{\rm o}}^{x}$ and $\upsilon_{{\rm e}}^{x}$.
When couplings are drawn from disorder distributions sufficiently
close to the IRFP form of Eq.~(\ref{eq:P*J-Fisher}), $\upsilon_{{\rm o}}^{x}$
and $\upsilon_{{\rm e}}^{x}$ are expected to approach the constant
values $\upsilon_{{\rm o}}^{x}=-1/4$ and $\upsilon_{{\rm e}}^{x}=0$.

\begin{figure}
\begin{center}\includegraphics[%
  clip,
  width=0.99\columnwidth,
  keepaspectratio]{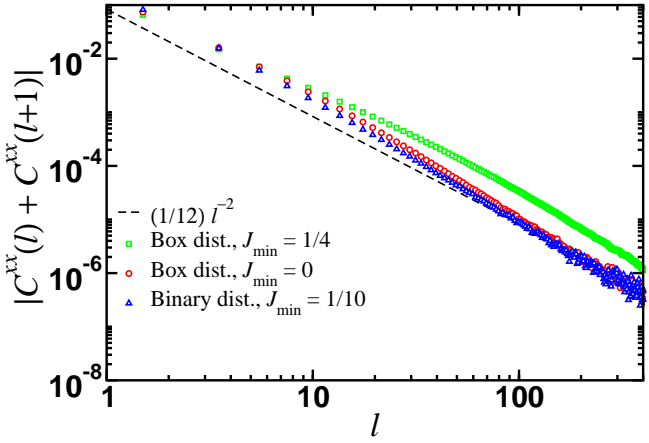}\end{center}

\caption{\label{cap:sum-transverse-C}(Color online) The summed transverse
correlation function $C_{{\rm sum}}^{xx}\left(l\right)=C^{xx}(l)+C^{xx}(l+1)$
as a function of the distance $l$ (for odd $l$) for XX chains and
the same coupling distributions as in the previous figure. Although
for sufficiently strong disorder (circles and triangles) the curves
approach a power law, corresponding to the random-singlet exponent
$\eta=2$ and to a prefactor $\left(\upsilon_{{\rm o}}^{x}+\upsilon_{{\rm e}}^{x}\right)/3\simeq-1/12$,
the less disordered system (squares) shows a different prefactor.
Again, results for $l>400$ and error bars are omitted for clarity.}
\end{figure}

The results from different coupling distributions provide evidence
that $\upsilon_{{\rm o}}^{x}$ and $\upsilon_{{\rm e}}^{x}$ indeed
approach constant values for arbitrary initial disorder, with $-\upsilon_{{\rm o}}^{x}$
and $\upsilon_{{\rm e}}^{x}$ assuming close (but certainly distinct)
values. Additionally, it seems that the quantity $\upsilon_{{\rm o}}^{x}+\upsilon_{{\rm e}}^{x}$
approaches an asymptotic value close to $-1/4$ for sufficiently strong
disorder. This can be seen in Fig.~\ref{cap:sum-transverse-C}, where
we plot (for $l$ odd) the combination $C_{{\rm sum}}^{xx}\left(l\right)\equiv-[C^{xx}(l)+C^{xx}(l+1)]$.
Notice that, for the box distribution with $J_{{\rm min}}=0$ and
the binary distribution with $J_{{\rm min}}=1/10$, the curves for
$C_{{\rm sum}}^{xx}(l)$ are reasonably well described by the scaling
form $1/\left(12l^{2}\right)$ in the long-distance limit. However,
this is not the case for chains with couplings drawn from the box
distribution with $J_{{\rm min}}=1/4$, at least up to the sizes studied
($l=1\,000$, not shown). Indeed, we argue in Sec.~\ref{sec:An-exactly-solvable}
that deviations from that scaling form should be expected for the
\emph{transverse} correlations in XX chains.

Finally, we report that we have considered also smaller chains ($1\,000$
sites) but with more disordered distributions ($J_{{\rm min}}=0,$
with $\vartheta_{0}=0.3$ or $\vartheta_{0}=0.6$). For the sake of
clarity, we have omitted their data in Figs.~\ref{cap:longitudinal-C}-\ref{cap:sum-transverse-C}.
The mean \emph{longitudinal} correlation function is remarkably well
described by the naive SDRG prediction (\ref{eq:corr-universal-aa}).
The mean \emph{transverse} correlation function, on the other hand,
is well described by Eq.~(\ref{eq:C-transversal}) with $\upsilon_{{\rm o}}^{x}+\upsilon_{{\rm e}}^{x}\approx-1/4$.

\subsection{XXX chains\label{sub:XXX-chains}}

\begin{figure}
\begin{center}\includegraphics[%
  clip,
  width=1\columnwidth,
  keepaspectratio]{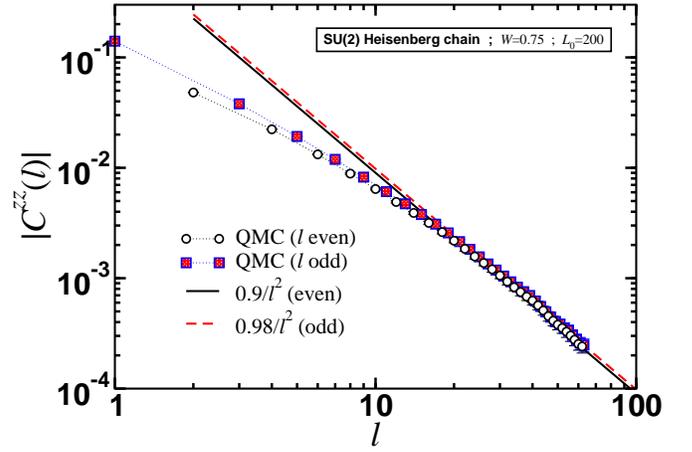}\end{center}

\caption{\label{fig:CzQMC}(Color online) Correlation function $(-1)^{l}C^{zz}(l)$
in the ground state of isotropic random AF Heisenberg spin-1/2 chains
{[}Eq.~(\ref{HamilXXX}){]} of length $L_{0}=200$ with disorder
strength $W=0.75$. The quantum Monte Carlo results were obtained
at $T/\overline{J}=1.5\times10^{-5}$ and averaged over $N_{{\rm samples}}=500$
realizations. When the distance $l$ between spins is even (circles),
the asymptotic regime is described by $C^{zz}(l)\simeq0.9/l^{2}$
(black, solid line) whereas for odd $l$ (squares), the best fit gives
$C^{zz}(l)\simeq-0.98/l^{2}$ (red, dashed line). }
\end{figure}

\begin{figure}
\begin{center}\includegraphics[%
  clip,
  width=1\columnwidth,
  keepaspectratio]{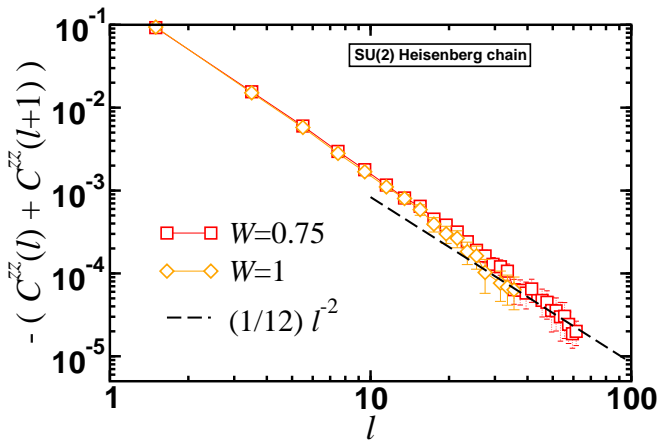}\end{center}

\caption{\label{fig:CzsumQMC}(Color online) Summed correlation function $C_{{\rm sum}}^{zz}(l)=|C^{zz}(l)+C^{zz}(l+1)|$
in the ground state of isotropic random AF Heisenberg spin-1/2 chains
{[}Eq.~(\ref{HamilXXX}){]} of length $L_{0}=200$ with disorder
strength $W=0.75$ (red squares) and $L_{0}=100$ with $W=1$ (orange
diamonds). The quantum Monte Carlo results were obtained at $T/\overline{J}=1.5\times10^{-5}$
for $W=0.75$ and $T/\overline{J}=4\times10^{-6}$ for $W=1$ and
averaged over $N_{{\rm samples}}=500$ realizations for each disorder
strength. The dashed line is the $1/\left(12l^{2}\right)$ prediction. }
\end{figure}

We now present QMC results obtained for the SU(2) symmetric model,
\begin{equation}
H=\sum_{i=1}^{L_{0}}J_{i}\mathbf{S}_{i}\cdot\mathbf{S}_{i+1}~,\label{HamilXXX}\end{equation}
 with the random AF couplings $J_{i}$'s distributed according to
the box distributions \begin{equation}
P\left(J\right)=\frac{1}{2\overline{J}W}\theta\left(J-\overline{J}\left(1-W\right)\right)\theta\left(\overline{J}\left(1+W\right)-J\right)~.\label{box}\end{equation}
 The QMC algorithm we use is based on a stochastic series expansion
of the partition function.~\cite{sandvik91,syljuasen-sandvik} This
is a finite temperature $T$ technique which, in principle, allows
access to ground-state properties, provided $T$ is chosen to be much
smaller than the finite-size gap of the system $\Omega\propto L_{0}^{-z}$.
As already discussed in several works (see, for instance, Refs.~\onlinecite{laflorencie-correlacao-PRB}
and \onlinecite{sandvik-perco2D,bergekvist-henelius-s1,laflorencie-RAF2D}),
the ground-state properties in random spin systems can be very hard
to access because extremely small energy correlations might develop
between distant spins or spin clusters. For random finite chains,
the dynamical exponent $z$ is formally infinite since we expect exponentially
small couplings to develop at large distances between spins, so that
$\Omega\propto\exp(-\sqrt{L_{0}})$. In order to accelerate the convergence
toward the ground state, we used the $\beta$-doubling scheme~\cite{sandvik-perco2D}
and thus performed the QMC measurements at temperatures as small as
$4\times10^{-6}$ in units of $\overline{J}$. We show in Figs.~\ref{fig:CzQMC}
and \ref{fig:CzsumQMC} QMC results for the average spin-spin correlation
function in the ground state, \begin{equation}
C^{zz}\left(l\right)=\frac{1}{N_{{\rm samples}}}\sum_{\sigma=1}^{N_{{\rm samples}}}\frac{2}{L_{0}}\sum_{i=1}^{L_{0}/2}\langle S_{i}^{z}S_{i+l}^{z}\rangle^{(\sigma)},\end{equation}
 where we perform disorder averaging over $N_{{\rm samples}}$ independent
random samples, as well as space averaging along the periodic chains.
Note that the SU(2) symmetry of the Hamiltonian ensures that $C^{zz}(l)=C^{yy}(l)=C^{xx}(l)$.

As studied in great detail in Refs.~\onlinecite{laflorencie-correlacao-PRL}
and \onlinecite{laflorencie-correlacao-PRB}, there is a crossover
phenomenon which is governed by the localization length $\xi$ of
the corresponding one-dimensional Jordan-Wigner fermions with random
hoppings. In order to be in the asymptotic regime, i.e., in the vicinity
of the IRFP, we have to look at system sizes $L_{0}\gg\xi$. For the
SU(2) symmetric case, $\xi$ has been estimated to be $\simeq\!20$
for $W=0.75$ and $\simeq\!10$ for $W=1$.~\cite{laflorencie-correlacao-PRB}
Thus, in order to study the IRFP asymptotic regime, we study two different
systems: $W=0.75$ with $L_{0}=200$ sites (see Fig.~\ref{fig:CzQMC})
and $W=1$ with $L_{0}=100$ (see Fig.~\ref{fig:CzsumQMC}). In Fig.~\ref{fig:CzQMC},
we first clearly see the crossover behavior for distances $l<20$
and then the IRFP prediction, Eq.~(\ref{eq:C-Fisher}), recovered
for larger separations. On the other hand, the prediction of Eq.~(\ref{eq:corr-universal-aa})
is not verified, and we confirm the observation already made for the
XX case. Again, we can write \begin{equation}
C^{\alpha\alpha}\left(l\gg\xi\right)=\frac{1}{3}\left\{ \begin{array}{ll}
\upsilon_{{\rm o}}l^{-2} & \textrm{if }l\textrm{ is odd,}\\
\upsilon_{{\rm e}}l^{-2} & \textrm{otherwise,}\end{array}\right.\label{eq:C-transversalXXX}\end{equation}
 where $\upsilon_{{\rm o}}$ and $\upsilon_{{\rm e}}$ are disorder-dependent
prefactors.

Nevertheless, the universality is recovered when looking at the sum
of the prefactors (see Fig.~\ref{fig:CzQMC}) $\upsilon_{{\rm o}}+\upsilon_{{\rm e}}\simeq-0.08\simeq-1/12$.
As shown in Fig.~\ref{fig:CzsumQMC}, the quantity \begin{equation}
C_{{\rm sum}}^{zz}\left(l\right)=-\left[C^{zz}\left(l\right)+C^{zz}\left(l+1\right)\right]\end{equation}
 seems to behave as $1/\left(12l^{2}\right)$ for $W=0.75$ and $W=1$.

\subsection{Discussion}

The origin of the apparent universality of $\upsilon_{{\rm o}}+\upsilon_{{\rm e}}=-1/4$
is not obvious. It is clear that the breakdown of the SDRG prediction
($\upsilon_{{\rm o}}=-1/4$, $\upsilon_{{\rm e}}=0$) must be related
to the fact that a collection of singlet pairs is not an exact eigenstate
of the Hamiltonian for any finite disorder. Although the SDRG method
becomes asymptotically exact at low energies, decimations involving
spins connected by couplings of the order of the initial energy scale
$\Omega_{0}$ unavoidably lead to significant errors due to the fact
that the calculation is perturbative. Thus, instead of singlet pairs,
these steps should really involve blocks of three or more neighboring
spins, so that correlations {}``spread'' over a few sites (whose
number decreases as the strength of the initial disorder increases),
forming clusters of correlated spins.

\begin{figure*}
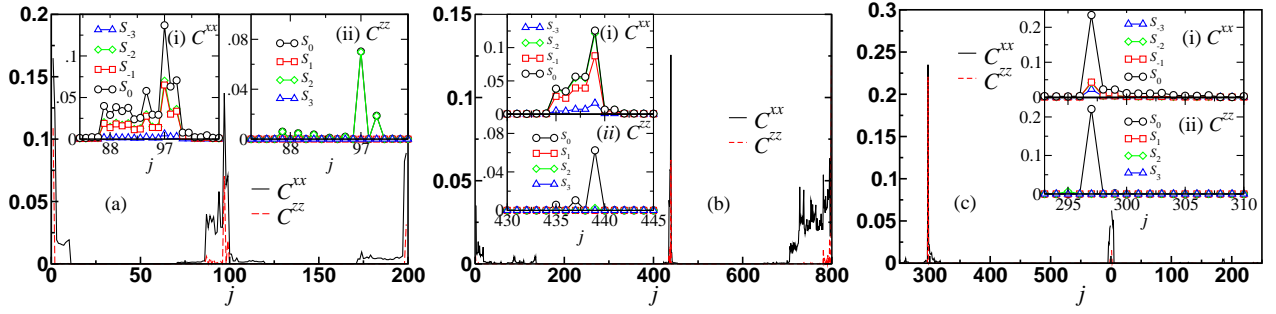

\begin{center}\includegraphics[%
  clip,
  width=0.31\textwidth,
  keepaspectratio]{fig8a.eps}~\includegraphics[%
  clip,
  width=0.31\textwidth,
  keepaspectratio]{fig8b.eps}~\includegraphics[%
  clip,
  width=0.31\textwidth,
  keepaspectratio]{fig8c.eps}\end{center}

\caption{\label{cap:Cxx-and-Czz}(Color online) The magnitude of the transverse
$C^{xx}$ and longitudinal $C^{zz}$ correlation functions between
$\mathbf{S}_{0}$ and $\mathbf{S}_{j}$ as a function of $j$ (for
$j\neq0$). Main panels (a), (b), and (c) correspond to different
samples drawn from different disorder distributions {[}see Eq.~(\ref{eq:box}){]}
whose parameters $(J_{{\rm min}},\vartheta_{0})$ are $(0.25,1)$,
$(0,1)$, and $(0,0.7)$, respectively. Insets (i) and (ii) show $C^{xx}$
and $C^{zz}$ between spins $\mathbf{S}_{i}$ and $\mathbf{S}_{j}$
as a function of $j$ for different values of $i$ between $-3$ and
$3$ and for values of $j$ around $j^{*}$. In the SDRG picture for
these particular realizations, $\mathbf{S}_{0}$ and $\mathbf{S}_{j^{*}}$
should couple in a singlet state where $j^{*}$ equals $97$, $439$,
and $297$ in panels (a), (b), and (c), respectively. The lines in
the insets are guides to the eyes.}
\end{figure*}

Figure \ref{cap:Cxx-and-Czz} shows the correlation functions $C_{j}^{xx}=\left\langle S_{0}^{x}S_{j}^{x}\right\rangle $
and $C_{j}^{zz}=\left\langle S_{0}^{z}S_{j}^{z}\right\rangle $ between
a reference spin $\mathbf{S}_{0}$ and the $j$-th neighboring spin
to the right $\mathbf{S}_{j}$ as a function of $j$ for XX chains
with couplings drawn from boxlike distributions {[}panels (a) and
(b){]} and a power-law distribution {[}panel (c){]}. For the particular
realizations in the figure, we find that, decimating the chains according
to the SDRG scheme, $\mathbf{S}_{0}$ should couple to $\mathbf{S}_{j^{*}}$
in a singlet pair, where $j^{*}=97$, $439$, and $297$ for panels
(a), (b), and (c), respectively. Indeed, there is a pronounced peak
at $j^{*}$, as expected from SDRG. In addition, $\mathbf{S}_{0}$
also develops strong correlations with a few spins adjacent to $\mathbf{S}_{j^{*}}$
and $\mathbf{S}_{0}$. As expected for a localized phase, these contributions
vanish exponentially at larger distances.

We now can define two spin clusters: The first one is composed by
$\mathbf{S}_{0}$ and its neighbors such that the magnitude of the
transverse correlation function between $\mathbf{S}_{0}$ and a spin
belonging to that cluster is bigger than a certain cutoff, say, $10^{-3}$,
the second cluster is analogous to the first one, but replacing $\mathbf{S}_{0}$
by $\mathbf{S}_{j^{*}}$. Interestingly, the sum of all the \emph{longitudinal}
correlations between the spins in the first cluster and the spins
in the second cluster approaches $-1/4$. We thus say that the correlation
between $\mathbf{S}_{0}$ and $\mathbf{S}_{j^{*}}$ {}``spreads''
among the spins in the clusters, and each of them acts collectively
as a single spin, leading to the universal result that $\upsilon_{{\rm o}}+\upsilon_{{\rm e}}=-1/4$
at large length scales. However, such feature is not verified when
we consider the \emph{transverse} correlation function. As we show
in the next section, this is related to the lack of total spin conservation
in the transverse direction.

\section{Exactly solvable model\label{sec:An-exactly-solvable}}

The discussion at the end of Sec.~\ref{sec:Numerical-results} suggests
that the formation of spin clusters is responsible for the failure
of the prediction of universal correlation functions, Eq.~(\ref{eq:corr-universal}).
According to the Marshall-Lieb-Mattis theorem,~\cite{marshall-prsla-55,lieb-mattis}
the ground state of a spin cluster with antiferromagnetic couplings
is a singlet, if the number of spins is even, or a doublet, if the
number of spins is odd. Thus, at low enough temperature, a cluster
with an even number of spins does not contribute to the magnetic properties
of the chain, while a cluster with an odd number of spins can be represented
by an effective spin-$1/2$ object.

To gain insight into the origin of the apparent universality of $\upsilon_{{\rm o}}+\upsilon_{{\rm e}}$
observed in the numerical calculations of Sec.~\ref{sec:Numerical-results},
we now consider a chain that, at low energies, can be interpreted
as being composed of a certain fraction $\epsilon$ of {}``effective''
spins and a fraction $1-\epsilon$ of remaining {}``original'' spins.
Each effective spin represents a cluster with an odd number of original
spins. For simplicity, we assume here that each effective spin represents
a cluster with only three original spins. In addition, we locate the
effective spin at the position corresponding to the middle spin of
the underlying three-spin cluster. With this restriction, if the effective
chain contains $\tilde{N}$ spins (original and effective ones), there
are, in fact, $N=\tilde{N}\left(1+2\epsilon\right)$ underlying original
spins involved. Moreover, if there are $\tilde{l}-1$ spins between
a given spin pair in the effective chain, we say that $\tilde{l}$
is the {}``effective'' distance between the spins in that pair.
{[}This is \emph{not} the same as the renormalized distance defined
in Eq.~(\ref{eq:l-tilde}).{]}

Now, we choose the couplings in the effective chain from a probability
distribution like that in Eq.~(\ref{eq:dist-inc}), with $\vartheta_{0}\ll1$,
so that it is sufficiently close to the infinite-disorder fixed-point
distribution and the occurrence of {}``bad'' decimations is highly
improbable. With this choice of couplings, it follows from Eq.~(\ref{eq:Ps})
that, in terms of the effective lengths $\tilde{l}$, the effective
singlet distribution is given by\begin{equation}
\tilde{P}_{{\rm s}}\left(\tilde{l}\right)=\frac{2}{3\tilde{l}^{2}}~,\end{equation}
 for odd $\tilde{l}$, while $\tilde{P}_{{\rm s}}\left(\tilde{l}\right)=0$
for even $\tilde{l}$. Therefore, the average number of singlets of
effective length $\tilde{l}$ in the effective chain is given by \begin{equation}
\tilde{N}_{{\rm s}}\left(\tilde{l}\right)=\frac{1}{2}\tilde{N}\tilde{P}_{{\rm s}}\left(\tilde{l}\right)~,\end{equation}
 with $\tilde{l}$ restricted to odd values.

\begin{figure}
\begin{center}\includegraphics[%
  clip,
  width=0.75\columnwidth,
  keepaspectratio]{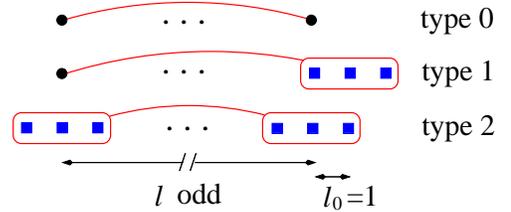}\end{center}

\caption{\label{cap:singlet-types}(Color online) The various types of singlets.
Black circles denote isolated original spins. Blue squares denote
original spins belonging to effective three-spin clusters. }
\end{figure}

In order to obtain the ground-state correlations in the underlying
chain, we have to determine the distribution of singlet lengths in
terms of the underlying original distances $l$, and these will depend
on how many effective spins are located between the two spins in a
given singlet. Let us consider a singlet formed between spins separated
by an effective distance $\tilde{l}$. Hence, there are $\tilde{l}-1$
intermediate spins, $m$ of which we assume are effective ones. Note
that there are three possible types of singlets (see Fig.~\ref{cap:singlet-types}):
a pair of original spins (type $0$), one original spin and one effective
spin (type $1$), and a pair of effective spins (type $2$). If the
singlet is of type $0$, then the underlying distance $l$ is given
by $l=\tilde{l}+2m$; if the singlet is of type $1$, then $l=\tilde{l}+2m+1$;
and for singlets of type $2$, $l=\tilde{l}+2m+2$. We immediately
conclude that, while singlets of types $0$ and $2$ are associated
with odd underlying distances $l$ (since $\tilde{l}$ is odd), singlets
of type $1$ are associated with even underlying distances. This leads
to the appearance of correlations between spins separated by even
distances, as in our numerical calculations, and in contrast to the
assumption of the SDRG approach.

For a singlet of length $\tilde{l}$, the number of intermediate effective
spins varies between $0$ and $\tilde{l}-1$. The probability of finding
exactly $m$ intermediate effective spins is given by\[
\left(\begin{array}{c}
\tilde{l}-1\\
m\end{array}\right)\epsilon^{m}\left(1-\epsilon\right)^{\tilde{l}-1-m}.\]
 Thus, the numbers of singlets of types $0$, $1$, and $2$ with
underlying length $l$ are given, respectively, by\begin{equation}
N_{{\rm s},0}\left(l\right)=\frac{1}{2}\tilde{N}\times\left(1-\epsilon\right)^{2}R\left(l\right)~,\end{equation}
\begin{equation}
N_{{\rm s},1}\left(l\right)=\frac{1}{2}\tilde{N}\times2\epsilon\left(1-\epsilon\right)R\left(l-1\right)~,\end{equation}
 and \begin{equation}
N_{{\rm s},2}\left(l\right)=\frac{1}{2}\tilde{N}\times\epsilon^{2}R\left(l-2\right)~,\end{equation}
 with\begin{equation}
R(l)=\frac{2}{3}\sum_{m=0}^{\frac{l-1}{3}}\left(\begin{array}{c}
l-2m-1\\
m\end{array}\right)\frac{\epsilon^{m}\left(1-\epsilon\right)^{l-3m-1}}{\left(l-2m\right)^{2}}~.\end{equation}
 In the limit of large $l$, the sum in $R(l)$ can be written as
an integral, which can be calculated by Laplace's method, using Stirling's
approximation. The final result is\begin{equation}
R(l)=\frac{2}{3}\left(1+2\epsilon\right)l^{-2}+\mathcal{O}\!\left(l^{-3}\right).\end{equation}
 Consequently,\begin{equation}
N_{{\rm s},0}\left(l\right)=\frac{N}{3}\left(1-\epsilon\right)^{2}l^{-2}+\mathcal{O}\!\left(l^{-3}\right)\qquad(\textrm{for odd $l$}),\label{eq:ns0}\end{equation}
\begin{equation}
N_{{\rm s},1}\left(l\right)=\frac{2N}{3}\epsilon\left(1-\epsilon\right)l^{-2}+\mathcal{O}\!\left(l^{-3}\right)\qquad(\textrm{for even $l$}),\label{eq:ns1}\end{equation}
 and \begin{equation}
N_{{\rm s},2}\left(l\right)=\frac{N}{3}\epsilon^{2}l^{-2}+\mathcal{O}\!\left(l^{-3}\right)\qquad(\textrm{for odd $l$}).\label{eq:ns2}\end{equation}

In order to calculate the ground-state correlations $C^{\alpha\alpha}\left(l\right)=\overline{\left\langle S_{i}^{\alpha}S_{i+l}^{\alpha}\right\rangle }$,
with $\alpha=x,\, y,\, z$, let us focus on a fixed odd value of $l$.
There are contributions to $C^{\alpha\alpha}\left(l\right)$ and $C^{\alpha\alpha}\left(l+1\right)$
coming from singlets of type $0$, with underlying length $l$; from
singlets of type $1$, with lengths $l-1$ and $l+1$; and from singlets
of type $2$, with lengths $l-2$, $l$, and $l+2$. If we denote
by $c_{\mathrm{end}}^{\alpha}$ ($c_{\mathrm{mid}}^{\alpha}$) the
{}``weight'' of a spin in either end (in the middle) of a three-spin
cluster to the $\alpha$ component of an effective spin (see Appendix~\ref{sec:Renormalization-of-a}),
we can combine all contributions to write (see Fig.~\ref{cap:singlet-types})\begin{align}
C^{\alpha\alpha}\left(l\right)= & -\frac{1}{4N}\left\{ N_{{\rm s},0}\left(l\right)+\left[N_{{\rm s},1}\left(l-1\right)+N_{{\rm s},1}\left(l+1\right)\right]c_{\mathrm{end}}^{\alpha}\phantom{\left(c_{\mathrm{mid}}^{\alpha}\right)^{2}}\right.\nonumber \\
 & +\left.\left[N_{{\rm s},2}\left(l-2\right)+N_{{\rm s},2}\left(l+2\right)\right]\left(c_{\mathrm{end}}^{\alpha}\right)^{2}\right.\nonumber \\
 & +\left.N_{{\rm s},2}\left(l\right)\left[2\left(c_{\mathrm{end}}^{\alpha}\right)^{2}+\left(c_{\mathrm{mid}}^{\alpha}\right)^{2}\right]\right\} ~,\label{eq:any}\end{align}
 and\begin{align}
C^{\alpha\alpha}\left(l+1\right)= & -\frac{1}{4N}\left\{ N_{{\rm s},1}\left(l+1\right)c_{\mathrm{mid}}^{\alpha}\right.\nonumber \\
 & +\left.2\left[N_{{\rm s},2}\left(l\right)+N_{{\rm s},2}\left(l+2\right)\right]c_{\mathrm{end}}^{\alpha}c_{\mathrm{mid}}^{\alpha}\right\} ~,\label{eq:any2}\end{align}
 so that, to leading order in $l$, we have\begin{align}
C^{\alpha\alpha}\left(l\right)= & -\frac{1}{12}l^{-2}\left\{ \left(1-\epsilon\right)^{2}+4\epsilon\left(1-\epsilon\right)c_{\mathrm{end}}^{\alpha}\phantom{\left(c_{\mathrm{mid}}^{\alpha}\right)^{2}}\right.\nonumber \\
 & +\left.\epsilon^{2}\left[4\left(c_{\mathrm{end}}^{\alpha}\right)^{2}+\left(c_{\mathrm{mid}}^{\alpha}\right)^{2}\right]\right\} \label{eq:any3}\end{align}
 and\begin{equation}
C^{\alpha\alpha}\left(l+1\right)=-\frac{1}{12}l^{-2}\left\{ 2\epsilon\left(1-\epsilon\right)+4\epsilon^{2}c_{\mathrm{end}}^{\alpha}\right\} c_{\mathrm{mid}}^{\alpha}.\end{equation}

Note that the above results are significantly different from the bare
SDRG results of Sec.~\ref{sec:Mean-corr-func}, most notably in that
the average correlation is, in general, not zero for even $l$. Both
$C^{\alpha\alpha}\left(l\right)$ and $C^{\alpha\alpha}\left(l+1\right)$
decay with the random-singlet exponent $\eta=2$, but with different
prefactors $\upsilon_{{\rm o}}^{\alpha}$ and $\upsilon_{{\rm e}}^{\alpha}$,
respectively. However, we have\begin{equation}
\frac{1}{3}\left(\upsilon_{{\rm o}}^{\alpha}+\upsilon_{{\rm e}}^{\alpha}\right)=-\frac{1}{12}\left\{ 1-\left[1-\left(2c_{\mathrm{end}}^{\alpha}+c_{\mathrm{mid}}^{\alpha}\right)\right]\epsilon\right\} ^{2}.\end{equation}

For the XXZ chain, irrespective of the initial anisotropy $\Delta$,
the $z$ component of the total spin is a good quantum number, assuming
the value $S_{{\rm tot}}^{z}=0$ in the (singlet) ground state. This
means that the sum of the ground-state correlations $\left\langle S_{i}^{z}S_{j}^{z}\right\rangle $
between a given spin $i$ and all other spins $j$ in the chain is
equal to $-1/4$. Since this is also true for an effective spin, it
follows that $2c_{\mathrm{end}}^{z}+c_{\mathrm{mid}}^{z}=1$ (as can
be easily verified explicitly; see Appendix \ref{sec:Renormalization-of-a}),
and we must have\begin{equation}
\upsilon_{{\rm o}}^{z}+\upsilon_{{\rm e}}^{z}=-\frac{1}{4}~,\end{equation}
 irrespective of the concentration $\epsilon$ of effective spins
(and thus of the initial disorder). Furthermore, in the Heisenberg
limit ($\Delta=1$), owing to the SU(2) symmetry, we also have\begin{equation}
\upsilon_{{\rm o}}^{x}+\upsilon_{{\rm e}}^{x}=\upsilon_{{\rm o}}^{y}+\upsilon_{{\rm e}}^{y}=\upsilon_{{\rm o}}^{z}+\upsilon_{{\rm e}}^{z}=-\frac{1}{4}~.\end{equation}
 This last result, however, is not valid for $\Delta<1$. In particular,
in the XX limit, for which $2c_{\mathrm{end}}^{x}+c_{\mathrm{mid}}^{x}\simeq0.9142$,
we obtain \begin{equation}
\upsilon_{{\rm o}}^{x}+\upsilon_{{\rm e}}^{x}\simeq-\frac{1}{4}\left(1-0.0858\epsilon\right)^{2},\end{equation}
 yielding a weak dependence on $\epsilon$.

For the isotropic Heisenberg chain and for $C^{zz}(l)$, the analytical
results derived in this section are in agreement with the numerical
results in Sec.~\ref{sec:Numerical-results}, strengthening the conjecture
of a universal behavior for sum of prefactors of the longitudinal
ground-state correlations in random XXZ chains. The presence of larger
effective-spin clusters (as typically occurs for weaker disorder;
see Fig.~\ref{cap:Cxx-and-Czz}) should not change the conclusions
of this section concerning the universality of $\upsilon_{{\rm o}}^{z}+\upsilon_{{\rm e}}^{z}$,
since the sum of all the weights of the original spins belonging to
an effective cluster is identically 1 when computed with respect to
the symmetry axis.

\section{Entanglement entropy and its relation to the correlation function\label{sec:Entanglement-entropy}}

In this section, we discuss the relation between the entanglement
entropy $S\left(l\right)$ and the ground-state mean correlation function
$C\left(l\right)$.

The entanglement entropy between two complementary subsystems $A$
and $B$ is given by\begin{equation}
S\left(l\right)=-\textrm{Tr}\,\rho_{A}\,\ln\,\rho_{A}=-\textrm{Tr}\,\rho_{B}\,\ln\,\rho_{B}~,\label{eq:S-definition}\end{equation}
 where $l$ is the length of one of the subsystems, \begin{equation}
\rho_{A}=\textrm{Tr}_{B}\,\rho=\sum_{i}\left\langle \phi_{B}^{i}\right|\rho\left|\phi_{B}^{i}\right\rangle \end{equation}
 is the reduced density matrix obtained by tracing out the degrees
of freedom of subsystem $B$ in the ground-state density matrix $\rho=\left|\phi\right\rangle \left\langle \phi\right|$,
and $\left\{ \left|\phi_{B}^{i}\right\rangle \right\} $ is a set
of states spanning the degrees of freedom of $B$ (with a similar
definition for $\rho_{B}$).

\begin{figure}
\begin{center}\includegraphics[%
  clip,
  width=1\columnwidth,
  keepaspectratio]{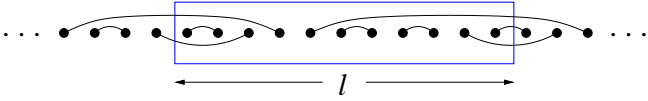}\end{center}

\caption{\label{cap:Ground-state}(Color online) Ground state of the infinite-disordered
AF spin-1/2 chain. The entanglement entropy between the subsystem
inside the box of length $l$ and the rest of the chain is equal to
the number of singlets shared by them. In this case, $S\left(l\right)=5\times s_{0}$,
with $s_{0}$ being the entanglement entropy of a singlet pair when
one of the spins is traced out.}
\end{figure}

In the SDRG framework, the ground state of random XXZ chains is a
collection of independent singlet pairs, i.e., \begin{equation}
\left|\phi\right\rangle =\bigotimes_{i=1}^{L_{0}/2}\left|0_{i}\right\rangle ~,\end{equation}
 where $\left|0_{i}\right\rangle $ denotes the $i$-th singlet pair
and $L_{0}$ is the total number of spins in the chain (see Fig.~\ref{cap:Ground-state}).

As the entanglement entropy between two spins in a singlet state is
$s_{0}=\ln\,2$,~\cite{foot-entropy} the total entanglement entropy
due to a given choice of subsystems $A$ and $B$ is equal to $s_{0}$
times the number of singlet pairs in which one spin belongs to $A$
and the other one to $B$ (see Fig.~\ref{cap:Ground-state}). Using
this fact, Refael and Moore~\cite{refael-moore} calculated the mean
number of times that each bond is decimated, which is equivalent to
the mean number of singlet lines crossing a given boundary. They found
that the mean value of the entanglement entropy grows as $\left(\gamma\,\ln\, l\right)/3$,
with $\gamma=\ln\,2$ being a universal number. This is reminiscent
of the entanglement entropy in conformally invariant (clean) one-dimensional
quantum systems, which increases as $\left(c\,\ln\, l\right)/3$,
where $c$ is the central charge, a signature of the universality
class of conformally invariant systems.~\cite{holzhey-larsen-wilczek,calabrese-cardy-jstatmech04}
In the clean critical XXZ chain, $c=1$.

\begin{figure}
\begin{center}\includegraphics[%
  clip,
  width=1\columnwidth,
  keepaspectratio]{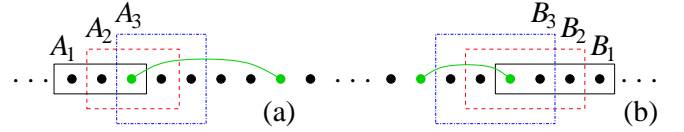}\end{center}

\caption{\label{cap:entropy-counting}(Color online) Schematic entropy counting.
(a) When the singlet length (in this case $l_{{\rm s}}=5$) is greater
than the subsystem length ($l_{A}=3$), there are $l_{A}$ different
subsystems whose right boundaries are crossed by the singlet. (b)
Otherwise, when the subsystem length $l_{B}$ is greater than $l_{{\rm s}}$
(in this case, $l_{B}=4$ and $l_{{\rm s}}=3$), there are $l_{{\rm s}}$
different subsystems whose left boundaries are crossed by the singlet.}
\end{figure}

We now rederive the mean entanglement entropy by relating it to the
distribution of singlet lengths {[}see Eq.~(\ref{eq:Ps}){]} and
thus to the SDRG mean correlation function {[}see Eq.~(\ref{eq:corr-universal}){]}.
By definition, the mean value of the entanglement entropy $S\left(l\right)$
between a subsystem of length $l$ and the rest of the chain is the
sum of the entropies of all subsystems of length $l$, divided by
$L_{0}$. (In a $L_{0}$-site chain with periodic boundary conditions,
there are $L_{0}$ different subsystems with the same length.) The
contribution of a given singlet of length $l_{{\rm s}}$ depends on
the relation between $l_{{\rm s}}$ and $l$. If $l_{{\rm s}}>l$,
there are $2l$ different subsystems of length $l$ whose boundaries
are crossed by the singlet line {[}see Fig.~\ref{cap:entropy-counting}(a){]}.
Likewise, $2l_{{\textrm{s}}}$ different subsystems have their boundaries
crossed by a singlet of length $l_{{\rm s}}\leq l$ {[}see Fig.~\ref{cap:entropy-counting}(b){]}.
Thus,\begin{equation}
S\left(l\right)=\frac{2s_{0}}{L_{0}}\left\{ \sum_{l_{{\rm s}}=1}^{l}l_{{\rm s}}N_{{\rm s}}\left(l_{{\rm s}}\right)+l\sum_{l_{{\rm s}}=l+1}^{L_{0}/2}N_{{\rm s}}\left(l_{{\rm s}}\right)\right\} ~,\label{eq:sumS-Ns}\end{equation}
 where $N_{{\rm s}}\left(l_{{\rm s}}\right)$ is the number of singlets
with length $l_{{\rm s}}$ in the ground state and $s_{0}=\ln\,2$
is the contribution of a singlet pair to the entanglement entropy.

As shown in Sec.~\ref{sec:Mean-corr-func}, $N_{{\rm s}}\left(l_{{\rm s}}\right)$
is simply related to the correlation function by $N_{{\rm s}}\left(l_{{\rm s}}\right)=-4L_{0}C\left(l_{{\rm s}}\right)/3$.
Thus, substituting Eq.~(\ref{eq:corr-universal}) into Eq.~(\ref{eq:sumS-Ns}),
and noting that $C\left(l_{{\rm s}}\right)=0$ for even $l_{{\rm s}}$,
we obtain, for $l\gg1$ and $L_{0}\rightarrow\infty$, \begin{align}
S\left(l\right)= & -\frac{8}{3}s_{0}\left\{ \sum_{l_{s}=1}^{l}l_{{\rm s}}C\left(l_{{\rm s}}\right)+l\sum_{l_{{\rm s}}=l+1}^{L_{0}/2}C\left(l_{{\rm s}}\right)\right\} \label{eq:mean-S}\\
= & \frac{2}{3}s_{0}\left(\frac{1}{2}\int_{\frac{2}{l}}^{1-\frac{1}{l}}\frac{1}{x}{\rm d}x+\frac{1}{2}l\int_{1+\frac{1}{l}}^{\infty}\frac{1}{x^{2}}{\rm d}x\right)+b^{'}\\
= & \frac{\gamma}{3}\ln\, l+b~,\label{eq:mean-S-2}\end{align}
 in which $\gamma=s_{0}=\ln\,2$, while $b$ and $b^{'}$ are nonuniversal
constants that depend on the short-distance details of $C\left(l\right)$.
In this way, we recover the result obtained by Refael and Moore. Yet
another derivation of the above result is presented in Appendix~\ref{sec:Another-derivation-of}.

Note that Eq.~(\ref{eq:mean-S}) relates the mean value of the entanglement
entropy to the mean correlation function. This relation is valid only
in the context of infinite-randomness spin chains, where both quantities
are dominated by rare spin singlets. In AF spin-1/2 chains without
disorder, for instance, such relation is no longer valid, and the
correct expression is far from simple (though an efficient valence-bond
approach can be developed to study block entanglement properties~\cite{alet-etal-07}).

\begin{figure}
\begin{center}\includegraphics[%
  clip,
  width=1\columnwidth,
  keepaspectratio]{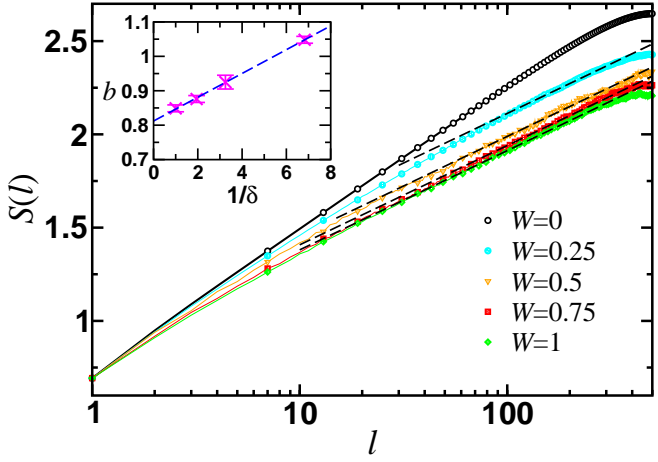}\end{center}

\caption{\label{fig:SXX}(Color online) Entanglement entropy of random XX
chains with $L_{0}=1\,000$ sites, and disorder of the form of Eq.~(\ref{box})
with, from top to bottom, $W=0$ (no disorder, open circles), $W=0.25$
(cyan circles), $W=0.5$ (orange triangles), $W=0.75$ (red squares),
and $W=1$ (green diamonds). These are exact diagonalization results
averaged over $10^{3}\leq N_{{\rm samples}}\leq10^{4}$ realizations.
The black dashed lines are fits of the form $S(l)=\left(1/3\right)\times\ln\,2\times\ln\, l+b$
with $b=1.048,~0.925,~0.876,~0.849$ for $W=0.25,~0.5,~0.75,~1$.
In the inset, the constant $b$ is also plotted versus $\delta^{-1}$,
where $\delta^{2}=\overline{(\ln J)^{2}}-\left(\overline{\ln J}\right)^{2}$,
and naively fitted to $b(\delta)=0.812+0.034\,56/\delta$.}
\end{figure}

Contrary to the naive universal form {[}Eq.~(\ref{eq:corr-universal-aa}){]}
of the ground-state correlation function, which is found not to hold
when confronted with exact diagonalization or QMC calculations, the
universal prediction of Eq.~(\ref{eq:mean-S-2}) is fully supported
by numerical results (see Ref.~\onlinecite{laflorencie-entanglement})
and, as shown in Fig.~\ref{fig:SXX}, does not depend on the initial
disorder strength. In view of the relation between these two quantities,
revealed by Eq.~(\ref{eq:mean-S}), the arising question is how these
seemingly contradictory results can be reconciled.

We address this question by looking at the entanglement entropy of
the exactly solvable model of Sec.~\ref{sec:An-exactly-solvable}.
The ground state of the model can be viewed as a collection of singlets
of three different types (see Fig.~\ref{cap:singlet-types}). So,
although Eq.~(\ref{eq:mean-S}) can no longer be used, the entanglement
entropy is still related to the distributions of singlet lengths,
in analogy with Eq.~(\ref{eq:sumS-Ns}). However, we must remember
that an effective spin represents a three-spin cluster. It can be
easily shown that the entanglement entropies between a three-spin
cluster and a single spin, as well as between two three-spin clusters,
are also given by $s_{0}=\ln\,2$. However, since we have to average
over all different possible subsystems of a given size, we must take
into account situations in which one of the spins in a three-spin
cluster lies in a different subsystem than the other two. When averaged
over all subsystems and singlet types, these {}``internal'' contributions,
being only \emph{boundary} effects, lead to an additional constant
term and thus do not affect the scaling properties of the entanglement
entropy. Explicitly, we have \begin{equation}
S\left(l\right)=\frac{1}{L_{0}}\left\{ S_{0}\left(l\right)+S_{1}\left(l\right)+S_{2}\left(l\right)\right\} ~,\label{eq:sleps}\end{equation}
 in which $S_{0}\left(l\right)$, $S_{1}\left(l\right)$, and $S_{2}\left(l\right)$
are the average entanglement entropies due to singlets of types $0$,
$1$, and $2$, respectively. From Eq.~(\ref{eq:sumS-Ns}), taking
into account the {}``internal'' contributions $b_{j}\left(\epsilon\right)$,
we can immediately write $S_{j}\left(l\right)$, for $j=0,$ $1$,
and $2$, as\begin{equation}
S_{j}(l)=b_{j}\left(\epsilon\right)+2s_{0}\left\{ \sum_{l_{{\rm s}}=1}^{l}l_{{\rm s}}N_{{\rm s},j}\left(l_{{\rm s}}\right)+l\sum_{l_{{\rm s}}=l+1}^{L_{0}/2}N_{{\rm s},j}\left(l_{{\rm s}}\right)\right\} ~,\end{equation}
 with $N_{{\rm s},j}\left(l\right)$ given by Eqs.~(\ref{eq:ns0})-(\ref{eq:ns2}).
Bearing in mind that $N_{{\rm s},0}\left(l\right)$ and $N_{{\rm s},2}\left(l\right)$
are zero for even $l$, while $N_{{\rm s},1}\left(l\right)$ is zero
for odd $l$, we can use the fact that\begin{equation}
N_{{\rm s},0}\left(l_{{\rm s}}\right)+N_{{\rm s},1}\left(l_{{\rm s}}-1\right)+N_{{\rm s},2}\left(l_{{\rm s}}\right)=\frac{L_{0}}{3}l_{{\rm s}}^{-2}+\mathcal{O}\!\left(l_{{\rm s}}^{-3}\right)~,\end{equation}
 for odd $l_{{\textrm{s}}}$, to conclude from Eq.~(\ref{eq:sleps})
that\begin{equation}
S\left(l\right)=\frac{\gamma}{3}\ln\, l+b\left(\epsilon\right)~,\end{equation}
 again with $\gamma=s_{0}=\ln\,2$, and an $\epsilon$-dependent constant
$b\left(\epsilon\right)$, as in Eq.~(\ref{eq:mean-S-2}).

Although this exactly solvable model yields a nonuniversal ground-state
correlation function $C\left(l\right)$, the entanglement entropy
$S\left(l\right)$ \emph{does} follow the universal form derived by
Refael and Moore.~\cite{refael-moore} This last result and the numerical
confirmation of the universality of $S\left(l\right)$ (see Ref.~\onlinecite{laflorencie-entanglement}
and Fig.~\ref{fig:SXX}) suggest that a description of the ground
state of random XXZ chains in terms of a collection of independent
singlets remains valid at sufficiently large distances, provided we
use the notion of effective spins already discussed in the previous
sections.

Finally, we should mention that the notion of the nonuniversality
of the correlation amplitude due to high-energy small-scale details
was first considered and investigated by Fisher and Young.~\cite{fisher-young-RTFIM}
Later, Refael and Moore~\cite{refael-moore} realized that such details
only contribute to the inaccuracies in the location of the low-energy
effective spins, which lead only to a surface term contribution to
the entanglement entropy, as we have formally shown here.

\section{Structure factor\label{sec:The-structure-factor}}

In this Section, we compute the \emph{static} structure factor \begin{align}
{\cal S}^{\alpha}\left(q\right)= & \frac{2\pi}{L_{0}}\sum_{j,k=1}^{L_{0}}e^{-iq\left(j-k\right)/l_{0}}\left\langle S_{j}^{\alpha}S_{k}^{\alpha}\right\rangle \nonumber \\
= & 2\pi\sum_{l=0}^{L_{0}-1}e^{-iql/l_{0}}C^{\alpha\alpha}\left(l\right),\label{eq:struc-factor}\end{align}
 which is straightforwardly related to the mean spin-spin time-independent
correlation function $C^{\alpha\alpha}\left(l\right)$ and is directly
measured in neutron scattering experiments. Indeed, neutron scattering
experiments probe the dynamical structure factor \begin{equation}
{\cal S}^{\alpha}\left(\omega,q\right)=\frac{1}{L_{0}}\sum_{j,k=1}^{L_{0}}e^{-iq\left(j-k\right)/l_{0}}\int_{-\infty}^{\infty}dte^{i\omega t}\left\langle S_{j}^{\alpha}\left(t\right)S_{k}^{\alpha}\left(0\right)\right\rangle ~,\end{equation}
 which reduces to ${\cal S}^{\alpha}\left(q\right)$ in Eq.~(\ref{eq:struc-factor})
after an integration over $\omega$.

There are three noteworthy properties of ${\cal S}^{\alpha}\left(q\right)$:
\begin{equation}
{\cal S}^{\alpha}\left(q\right)={\cal S}^{\alpha}\left(-q\right)~,\label{eq:property-1}\end{equation}
\begin{equation}
\sum_{q\in BZ}{\cal S}^{\alpha}\left(q\right)=\frac{1}{2}\pi L_{0}~,\label{eq:property-2}\end{equation}
 where the sum is over the first Brillouin zone, and \begin{equation}
{\cal S}^{\alpha}\left(q=0\right)=\frac{2\pi}{L_{0}}\left\langle \left(S_{{\rm tot}}^{\alpha}\right)^{2}\right\rangle ~,\label{eq:property-3}\end{equation}
 where $S_{{\rm tot}}^{\alpha}$ is the total spin along the $\alpha$
direction and $\left\langle \cdots\right\rangle $ means its expectation
value on the ground state. Hence, ${\cal S}^{z}\left(0\right)=0$
for the XX model and ${\cal S}^{x,y,z}\left(0\right)=0$ for the isotropic
case. Note that, in the continuum limit, Eq.~(\ref{eq:property-2})
leads to $\int_{-\pi}^{\pi}dq{\cal S}^{\alpha}\left(q\right)=\pi^{2}$.

We now show our numerical results on the static structure factor for
the disordered chain in the XX and XXX models. We anticipate that
all three properties {[}Eqs.~(\ref{eq:property-1})-(\ref{eq:property-3}){]}
are obeyed by our numerical results.

\subsection{XX model}

\begin{figure}
\begin{center}\includegraphics[%
  clip,
  width=1\columnwidth,
  keepaspectratio]{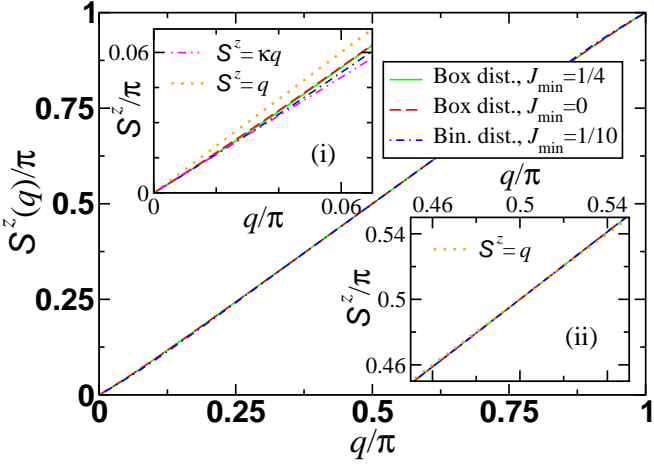}\end{center}

\caption{\label{cap:struc-factorZ-XX}(Color online) The longitudinal structure
factor in the XX model for various disordered chains of lengths up
to $L_{0}=4\,000.$ As can be seen, they are practically indistinguishable.
Inset (i) highlights ${\cal S}^{z}$ for small $q$ in which the curves
become somewhat distinguishable near $q=0.06\pi$. Moreover, they
slightly deviate from the clean system prediction ${\cal S}_{{\rm c}}^{z}=q$
(dotted line) tending to a \emph{universal} form ${\cal S}^{z}=\kappa\left|q\right|$
(dashed line), with $\kappa=\pi^{2}/12$ (see text). Inset (ii) shows
${\cal S}^{z}$ near $q=\pi/2$, where disorder is irrelevant. See
Eqs.~(\ref{eq:box}) and (\ref{eq:binary}) for the definition of
the disorder parameters (here $\Omega_{0}=1$ and $\vartheta_{0}=0$).}
\end{figure}

Figure~\ref{cap:struc-factorZ-XX} shows the longitudinal structure
factor in the XX model for the clean system {[}in which ${\cal S}_{{\rm c}}^{z}\left(q\right)=\left|q\right|$
(see inset){]} and various disordered chains. Because the longitudinal
mean correlation function depends very weakly on the disorder (only
through the crossover length) and its universal amplitude in the disordered
case is very close to the clean-system value ($1/12$ in comparison
to $1/\pi^{2}$; see Secs.~\ref{sub:The-mean-correlation} and \ref{sub:XX-chains}),
the longitudinal structure factor is, for all practical purposes,
universal.

In general, due to spin conservation and the fact that $C^{zz}\left(l\right)=0$
for even $l$, ${\cal S}^{z}$ equals $0$ and $\pi$ at the points
$q=0$ and $q=\pi$, respectively. Moreover, it can be shown that
${\cal S}^{z}$ has inversion symmetry around the point $q=\pi/2$,
i.e., ${\cal S}^{z}\left(\pi/2+k\right)+{\cal S}^{z}\left(\pi/2-k\right)=\pi$,
for $-\pi/2<k<\pi/2$, which implies ${\cal S}^{z}\left(\pi/2\right)=\pi/2$.

Let us now consider the effects of disorder more closely. Because
the behavior of ${\cal S}^{z}$ for small $q$ is dominated by the
large-$l$ behavior of the longitudinal correlation function, it follows
that ${\cal S}^{z}\left(q\ll1\right)\rightarrow\kappa\left|q\right|$,
where $\kappa=\pi^{2}\upsilon_{{\rm o}}^{z}/3=\pi^{2}/12\approx0.822$
is a \emph{universal} constant {[}see inset (i) of Fig.~\ref{cap:struc-factorZ-XX}{]}.
In the same way, because $C^{zz}\left(l\right)=0$ for even $l$,
the behavior near $q=\pi$ is also dominated by disorder. In this
case, $\pi-{\cal S}^{z}\left(q\approx\pi\right)\rightarrow\kappa\left|\pi-q\right|$.
This is verified by the data of Fig.~\ref{cap:struc-factorZ-XX},
but not shown for clarity. Finally, because the Fourier series {[}Eq.~(\ref{eq:struc-factor}){]}
at $q$ near $\pi/2$ selects small values of $l$, the behavior near
$q=\pi/2$ is dominated by the clean-system prediction, ${\cal S}^{z}\left(q\approx\pi/2\right)\rightarrow\left|q\right|$
as shown in inset (ii) of Fig.~\ref{cap:struc-factorZ-XX}. Note
that all these arguments are valid because $C^{zz}\left(l\right)=0$
for even $l$.

\begin{figure}
\begin{center}\includegraphics[%
  clip,
  width=1\columnwidth,
  keepaspectratio]{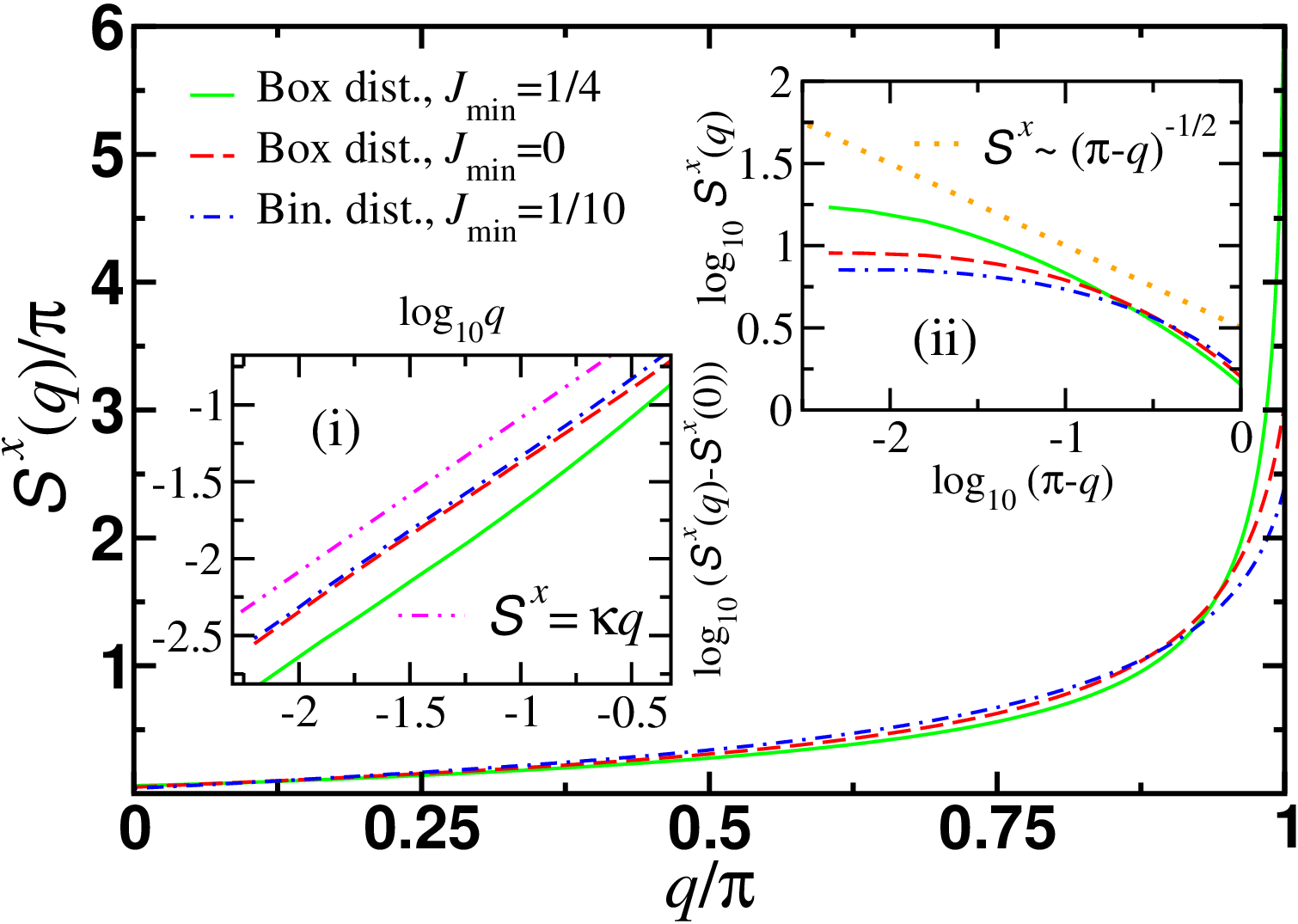}\end{center}

\caption{\label{cap:struc-factorX-XX}(Color online) The transverse structure
factor in the XX model for various disordered chains of lengths up
to $L_{0}=4,000.$ Inset (i) highlights ${\cal S}^{z}$ for small
$q$, where ${\cal S}^{x}\left(0\right)\approx0.194,~0.159,$ and
$0.128$ for the box distributions with $J_{{\rm min}}=1/4$, $J_{{\rm min}}=0$,
and the binary distribution, respectively. In all cases, ${\cal S}^{x}\left(q\right)-{\cal S}^{x}\left(0\right)\sim\left|q\right|$
. Inset (ii) shows the behavior of ${\cal S}^{x}$ near $q=\pi$.
It follows the characteristic divergence of the clean system prediction
(dotted line) until a crossover vector $q_{{\rm c}}$ above which
it saturates to a constant (see text).}
\end{figure}

We now discuss the behavior of ${\cal S}^{x}$ (see Fig.~\ref{cap:struc-factorX-XX}),
in which disorder plays a more prominent role. In the absence of disorder,
$C_{{\rm c}}^{xx}\left(l\right)\approx-\tilde{F}/l^{2K+1/\left(2K\right)}+\left(-1\right)^{l}F/l^{1/\left(2K\right)}$,
where $K=1$ is the Luttinger liquid parameter {[}see Eq.~(\ref{eq:cx-clean}){]},
and $\tilde{F}$ and $F\approx0.147\,09$ (Ref. \onlinecite{mccoy-68})
are positive constants. Since the first term is monotonic, it dominates
the structure factor for $q\ll1$. Hence, ${\cal S}_{{\rm c}}^{x}\left(q\ll1\right)-{\cal S}_{{\rm c}}^{x}\left(q=0\right)\sim\left|q\right|^{3/2}$.
The second (staggered) term gives a subdominant contribution $\sim q^{2}$.
Near the AF peak, $q=\pi$; however, the second term dominates, yielding
a divergent contribution, namely, ${\cal S}_{{\rm c}}^{x}\left(q=\pi-\epsilon\right)\sim\left|\epsilon\right|^{-1/2}$,
for $\left|\epsilon\right|\ll1$.

Quenched disorder dramatically changes the above scenario. Rewriting
the transverse correlation function {[}see Eq.~(\ref{eq:C-transversal}){]}
as $C^{xx}\left(l\gg1\right)\sim\left(\upsilon_{{\rm o}}^{x}+\upsilon_{{\rm e}}^{x}\right)\delta_{l,{\rm odd}}/\left(3l^{2}\right)+\left(-1\right)^{l}\upsilon_{{\rm e}}^{x}/\left(3l^{2}\right)$
(where $\delta_{l,{\rm odd}}=1$, if $l$ is odd, and $\delta_{l,{\rm odd}}=0$,
otherwise), it becomes clear that ${\cal S}^{x}\left(q\ll1\right)$
is dominated by the first term: ${\cal S}^{x}\left(q\ll1\right)-{\cal S}^{x}\left(q=0\right)\rightarrow-\pi^{2}\left(\upsilon_{{\rm o}}^{x}+\upsilon_{{\rm e}}^{x}\right)\left|q\right|/3$,
where $-\left(\upsilon_{{\rm o}}^{x}+\upsilon_{{\rm e}}^{x}\right)=1/4$
+ nonuniversal contributions {[}see inset (i) of Fig.~\ref{cap:struc-factorX-XX}{]}.
Note that this result strongly relies on the fact that $\upsilon_{{\rm o}}^{x}+\upsilon_{{\rm e}}^{x}$
is nonzero; otherwise, ${\cal S}^{x}\left(q\ll1\right)$ would be
dominated by the short length scale contributions to $C^{xx}$, which
follow the clean-system prediction.

The characteristic AF divergence near $q=\pi$ is suppressed by disorder,
as shown in inset (ii) of Fig.~\ref{cap:struc-factorX-XX}. For $q>q_{c}=2\pi/L_{c}$,
where $L_{c}$ is the crossover length above which the correlation
function is dominated by the disorder, ${\cal S}^{x}$ saturates to
a constant approximately equal to $2\pi F\sum_{l=1}^{L_{c}}1/\sqrt{l}$.
Finally, because Eq.~(\ref{eq:property-2}) has to be satisfied,
the decrease of the AF peak is accompanied by its broadening as disorder
increases {[}see inset (ii) of Fig.~\ref{cap:struc-factorX-XX}{]}.

\subsection{XXX Model}

\begin{figure}
\begin{center}\includegraphics[%
  clip,
  width=1\columnwidth,
  keepaspectratio]{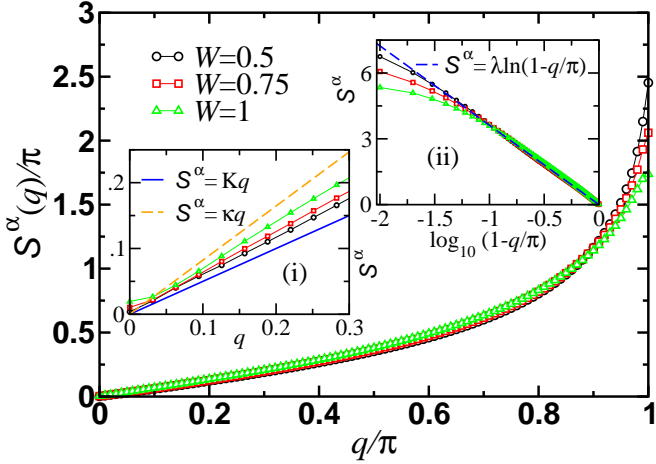}\end{center}

\caption{\label{cap:struc-factor-XXX}The structure factor as a function of
$q$ for different disorder parameters {[}see Eq.~(\ref{box}){]}
in the isotropic Heisenberg model for chains of lengths up to $L_{0}=200$.
Inset (i) shows ${\cal S}^{\alpha}$ for $q\ll1$ and compares it
with the (field-theoretical) clean system prediction ${\cal S}_{{\rm c}}^{\alpha}\rightarrow K\left|q\right|$,
with $K=1/2$ being the Luttinger liquid parameter, and the disordered
system prediction ${\cal S}^{\alpha}\rightarrow\kappa\left|q\right|$,
with universal $\kappa=\pi^{2}/12$. Inset (ii) highlights ${\cal S}^{\alpha}$
near the AF peak $q\approx\pi$. The dashed line is the clean-system
prediction, namely, ${\cal S}_{{\rm c}}^{\alpha}\rightarrow-\lambda\,\ln\left(1-q/\pi\right)$,
with $\lambda=-\pi/2$ (see text). }
\end{figure}

We now turn our attention to the isotropic Heisenberg model (see Fig.~\ref{cap:struc-factor-XXX}).

Similarly to the XX model, the universal features of the correlation
amplitude yield a universal structure factor for $q\ll1$: ${\cal S}^{\alpha}\left(q\right)\rightarrow\kappa\left|q\right|,$
$\forall\alpha$, since $\upsilon_{{\rm o}}+\upsilon_{{\rm e}}=1/4$
is universal. By coincidence, the clean system prediction also scales
linearly,~\cite{pereira-etal} ${\cal S}_{{\rm c}}^{\alpha}\left(q\ll1\right)\rightarrow K\left|q\right|$,
however, with a different slope $K=1/2$ which is the Luttinger liquid
parameter. As shown in inset (i) of Fig.~\ref{cap:struc-factor-XXX},
${\cal S}^{\alpha}\left(q\right)$ is linear, but apparently with
no universal slope. However, although we have statistical fluctuations
and finite-size effects, a close inspection of our data for $q\lesssim0.07$
shows that $\kappa\left|q\right|$ fits better than $K\left|q\right|$.

The logarithmic divergence at the $q=\pi$ point is suppressed by
disorder. As shown in inset (ii) of Fig.~\ref{cap:struc-factor-XXX},
${\cal S}^{\alpha}\left(q\right)$ follows the clean-system divergence
up to a crossover vector $q_{c}=2\pi/L_{c}$, where $L_{c}$ is the
crossover length above which the correlation exponent follows the
long-distance prediction of the disordered system. Beyond $q_{c}$,
${\cal S}^{\alpha}$ saturates at a non-universal constant proportional
to $2\pi\sum_{l=1}^{L_{c}}\sqrt{\ln\, l}/l$. We note that the clean-system
prediction depicted by the dashed line, ${\cal S}_{{\rm c}}^{\alpha}\rightarrow-\left(\pi/2\right)\ln\left(1-q/\pi\right)$,
is actually the prediction of the Haldane-Shastry model,~\cite{haldane-r-2-spinchain,shastry-longrange-spinchain}
which is a good approximation to the Heisenberg model for $q\lesssim13\pi/14$.~\cite{karbach-94,karbach-97}
For $q>13\pi/14$, ${\cal S}_{{\rm c}}^{\alpha}$ diverges as $[-\ln(1-q/\pi)]^{3/2}$,
consistent with $C_{{\rm c}}^{\alpha}\sim(-1)^{l}\sqrt{\ln\, l}/l$
for $l\gg1$.

\subsection{Discussion}

Summarizing, ${\cal S}^{\alpha}\left(q\right)$ is peaked at $q=\pi$,
$\forall\alpha$, in both models, while it approaches zero at $q=0$,
reflecting the antiferromagnetic quasi-long-range order. Near $q=\pi$,
the low-energy behavior of the structure factor is dominated by the
short-length scale behavior of the spin-spin correlation function,
and thus its scaling is determined by the physics of the clean system.
However, the true divergence is \emph{completely} suppressed by disorder,
and the peak becomes shorter and broader. On the other hand, the structure
factor vanishes \emph{universally} $\sim\left|q\right|$ for $q\ll1$
{[}for ${\cal S}^{x}$ in the XX model, one may consider the quantity
${\cal S}^{x}\left(q\right)-{\cal S}^{x}\left(q=0\right)${]} as a
consequence of two features: (i) the power-law scaling of the mean
spin-spin correlation function $C^{\alpha\alpha}\left(l\right)\sim l^{-\eta}$,
with universal exponent~\cite{doty-fisher,fisher94-xxz} $\eta=2$
\emph{and} (ii) the amplitudes $\upsilon_{{\rm o}}$ and $\upsilon_{{\rm e}}$
being different in magnitude. Moreover, due to the universal features
of these amplitudes, ${\cal S}^{\alpha}\left(q\right)-{\cal S}^{\alpha}\left(q=0\right)$
vanishes as $\kappa\left|q\right|$, with a universal $\kappa=\pi^{2}/12$,
if $\alpha$ is a symmetry axis, and a nonuniversal $\kappa$, otherwise.

We now briefly discuss a controversy that has appeared in the literature.
The dynamical structure factor ${\cal S}^{\alpha}\left(\omega,q\right)$
was theoretically studied in Refs. \onlinecite{huse-structurefactor-PRL}
and \onlinecite{huse-structurefactor-PRB} within the SDRG framework
(and thus, in the small-$\omega$ limit) and experimentally studied
in Refs.~\onlinecite{masuda,masuda-erratum,zheludev}, mainly by
measuring the local dynamical structure factor ${\cal S}\!\left(\omega\right)$
{[}obtained when one integrates ${\cal S}\!\left(\omega,q\right)$
over $q${]} for the compound BaCu$_{2}$(Si$_{0.5}$Ge$_{0.5}$)$_{2}$O$_{7}$.
Previously, it had been thought that this compound is a good experimental
realization of the random AF spin-1/2 chain with quenched bond randomness,
since both the experimentally determined static magnetic susceptibility
and local dynamical structure factor were found to be in good agreement
with the strong-disorder theoretical predictions.~\cite{masuda}
However, further and more precise measurements appeared to be in contradiction
with the strong-disorder theoretical scaling of ${\cal S}\!\left(\omega\right)$.~\cite{masuda-erratum,zheludev}
Interestingly, the magnetic susceptibility measurements remained in
agreement with the theoretical prediction for the disordered system.~\cite{masuda-erratum}
This led to a puzzle. Thermodynamical quantities seem to be dominated
by the physics of the disordered system, whereas spin correlations
seem to show the clean system physics.

We tentatively argue that the low-energy behavior of the quantity
$\omega\,{\cal S}\left(\omega\right)$ investigated experimentally
may \emph{not} be dominated by the physics of the disordered system,
even if the system itself is governed by a strongly disordered fixed
point. The quantity ${\cal S}\left(\omega\right)$ is obtained from
an integration over all values of $q$. Therefore, it is dominated
by the antiferromagnetic peak $q=\pi$. Such large momentum reflects
the shortest length scales of the time-dependent correlation function,
whose behavior is expected to be dominated by the physics of the disorder-free
system, as in the case of the time-independent correlation function.
Hence, the experimental determination of the full $q$-dependent ${\cal S}^{\alpha}\left(q,\omega\right)$
would be highly desirable.

Recently, it was shown~\cite{mucio-07} that the spin-1/2 compound
MgTiOBO$_{3}$ displays a remarkable random-singlet signature for
the magnetic susceptibility in a wide and accessible range of temperature.
It would be interesting to perform neutron scattering experiments
on this compound in order to check the predictions shown here for
the static structure factor.

\section{Conclusions\label{sec:Conclusions}}

In this paper, we revisited the ground-state properties of random-bond
antiferromagnetic quantum spin-$1/2$ chains using analytical and
numerical tools. We focused on the question of the universality of
the spin-spin correlation function $C\left(l\right)$ and of the average
entanglement entropy $S\left(l\right)$, as functions of the distance
$l$, as well as the connection between them.

By following exactly the SDRG flow of a family of coupling-constant
distributions, we showed that the SDRG approach predicts a fully universal
power-law scaling form of the pair correlation function $C\left(l\right)$,
in which both the prefactor and the decay exponent are disorder independent.
However, the SDRG prediction is strictly valid only in the limit of
infinite randomness. Exact diagonalization and quantum Monte Carlo
calculations on finite chains showed that this purported universality
does not hold, except for the correlations along the symmetry axis
in the XX model. Moreover, these numerical results reveal different
correlation amplitudes for spins separated by odd and even distances,
$\upsilon_{{\rm o}}$ and $\upsilon_{{\rm e}}$, respectively. Nevertheless,
we showed numerical evidence that the combination $\upsilon_{{\rm o}}+\upsilon_{{\rm e}}$,
at least for the XXX model and for correlations along the symmetry
axis in the XX model, is indeed universal and agrees with the SDRG
prediction if we consider that spin clusters themselves develop singletlike
correlations. In other words, the correlations of random singlets
spread among spins in the clusters.

As the average number of spins in a cluster depends on the details
of the coupling-constant distribution, correlation-function prefactors
are nonuniversal. However, the conservation law for the total spin
component along the symmetry axis guarantees that correlations {}``spread''
over all spins in the effective-spin singlets, leading to the universality
of $\upsilon_{{\rm o}}+\upsilon_{{\rm e}}=-1/4$. This hypothesis
was confirmed analytically in an exactly solvable model, in which
a number of three-spin clusters were introduced into an infinite-disorder
random-bond chain. Interestingly, the fact that $\upsilon_{{\rm o}}$
and $\upsilon_{{\rm e}}$ are different in magnitude has important
experimental relevance: the small-$q$ behavior of the structure factor
is dominated by disorder. We have argued that $q$-resolved neutron
scattering experiments may be able to probe the universal features
of those amplitudes.

We also rederived the average ground-state entanglement entropy $S\left(l\right)$,
relating it to the mean correlation function. In contrast to the nonuniversality
of $\upsilon_{{\rm o}}+\upsilon_{{\rm e}}$ when considering $C(l)$
along a nonsymmetry axis, the universal form of $S\left(l\right)$
first derived by Refael and Moore~\cite{refael-moore} was shown
to hold for the exactly solvable model with effective spins, in agreement
with the numerical data presented in Sec.~\ref{sec:Entanglement-entropy}
and previously elsewhere.~\cite{laflorencie-entanglement}

\emph{Note added}. Recently, we became aware of Ref.~\onlinecite{xu-ying-wan}
where the dynamical structure factor is also studied and the controversy
mentioned at the end of Sec.~\ref{sec:The-structure-factor} is considered. 

\begin{acknowledgments}
J.A.H. would like to thank R. G. Pereira, T. Vojta, and D. A. Huse
for useful discussions. J.A.H. and N.L. are grateful for the hospitality
of the Pacific Institute for Theoretical Physics and Les Houches Summer
School where part of this work was performed. This work was partially
supported by Fapesp under Grant No. 03/00777-3, by NSF under Grant
No. DMR-0339147, by Research Corporation (J.A.H.), by CNPq/FUNCAP
under Grant No. 350145/2005-9 (A.P.V.), by NSERC, by the Swiss National
Fund, by MaNEP (N.L.), and by CNPq under Grant No. 305971/2004-2 (E.M.).
Part of the simulations have been preformed using the WestGrid network. 
\end{acknowledgments}
\appendix

\section{Calculation of $P(\Omega,l)$\label{sec:Calculation-of-P}}

Laplace transforming Eq.~(\ref{eq:flow-joint-dist}) with respect
to the length variable $l$ yields \begin{equation}
\frac{\partial\hat{P}}{\partial\Omega}=-\hat{P}\left(\Omega,\lambda\right)\int{\rm d}J_{1}{\rm d}J_{3}\hat{P}\left(J_{1},\lambda\right)\hat{P}\left(J_{3},\lambda\right)\delta\left(J-\frac{J_{1}J_{3}}{\Omega}\right)~,\label{eq:flow-joint-dist-laplace}\end{equation}
 where $\hat{P}\left(J,\lambda\right)=\int\exp\left(-\lambda l\right)P\left(J,l\right){\rm d}l$.
We now substitute the \emph{Ansatz~}\cite{igloi-det-z-PRL,igloi-det-z-PRB}\begin{equation}
\hat{P}\left(J,\lambda\right)=\frac{\alpha\left(\lambda,\Omega\right)}{\Omega}\left(\frac{\Omega}{J}\right)^{1-\beta\left(\lambda,\Omega\right)}\end{equation}
 into Eq.~(\ref{eq:flow-joint-dist-laplace}) and find a pair of
equations,\begin{eqnarray}
\frac{{\rm d}}{{\rm d}\Gamma}\alpha & = & -\alpha\beta~,\\
\frac{{\rm d}}{{\rm d}\Gamma}\beta & = & -\alpha^{2}~,\end{eqnarray}
 with the boundary conditions $\alpha\left(\lambda=0,\Omega\right)=\beta\left(\lambda=0,\Omega\right)=\vartheta\left(\Omega\right)$.
Since \begin{equation}
\frac{{\rm d}}{{\rm d}\Gamma}\left(\beta^{2}-\alpha^{2}\right)=0~,\end{equation}
 we find the solutions\cite{igloi-det-z-PRB}\begin{eqnarray}
\beta & = & \frac{\beta_{0}\, c+c^{2}\tanh\left(c\Gamma\right)}{c+\beta_{0}\tanh\left(c\Gamma\right)}~,\label{eq:beta}\\
\alpha & = & \frac{c\sqrt{\beta_{0}^{2}-c^{2}}}{c\,\cosh\left(c\Gamma\right)+\beta_{0}\sinh\left(c\Gamma\right)}~,\label{eq:alfa}\end{eqnarray}
 where $c=c\left(\lambda\right)$ is a constant of the flow, defined
by $c^{2}=\beta^{2}-\alpha^{2}$. Moreover, $c$ is a real number
since $\beta>\alpha$, which can be shown by considering an explicit
calculation of the mean distance between the {}``active'' spins,\begin{eqnarray}
\bar{l} & = & \int{\rm d}l\,\, l\int{\rm d}JP\left(J,l\right)\nonumber \\
 & = & -\lim_{\lambda\rightarrow0}\frac{{\rm d}}{{\rm d}\lambda}\int_{0}^{\Omega}{\rm d}J\hat{P}\left(J,\lambda\right)\nonumber \\
 & = & \lim_{\lambda\rightarrow0}\frac{1}{\lambda}\left(1-\frac{\alpha\left(\lambda,\Omega\right)}{\beta\left(\lambda,\Omega\right)}\right)~.\label{eq:l-mean}\end{eqnarray}
 As $\bar{l}>0$, Eq.~(\ref{eq:l-mean}) ensures that $\alpha<\beta$.

The boundary conditions lead to $c\left(\lambda\rightarrow0\right)\ll1$,
and so $c^{2}\left(\lambda\ll1\right)=a^{2}\lambda^{\tau}+{\mathcal{O}}\!\left(\lambda^{\tau+1}\right)$.
From \begin{eqnarray}
\bar{l} & = & \lim_{\lambda\rightarrow0}\frac{1}{\lambda}\left(1-\frac{\sqrt{\beta^{2}-c^{2}}}{\beta}\right)\nonumber \\
 & = & \lim_{\lambda\rightarrow0}\frac{c^{2}}{2\lambda\beta^{2}}=\frac{a^{2}}{2\vartheta_{0}^{2}}\left(1+\vartheta_{0}\Gamma\right)^{2}\nonumber \\
 & = & \bar{l}_{0}\left(1+\vartheta_{0}\Gamma\right)^{2},\label{eq:l-medio-gama}\end{eqnarray}
 one finds $\tau=1$, the last step coming from the definition $\bar{l}=\bar{l}_{0}/n_{\Omega}$,
where $\bar{l}_{0}$ is the initial mean distance between the sites.
For simplicity, we consider that initially all spins are uniformly
separated, i.e., $l_{i}=\bar{l}_{0}=l_{0}$ is the lattice spacing.
The constant $a$ thus equals $\vartheta_{0}\sqrt{2l_{0}}$ and depends
on the two parameters of the problem, the initial length parameter
$l_{0}$ and the initial disorder parameter $\vartheta_{0}$. This
is a consequence of the fact that the magnitude of a coupling constant
$J$ shared by two spins a distance $l$ apart is correlated with
$l$, and thus the joint distribution $P\left(J,l\right)$ cannot
be written as $P_{J}\left(J\right)P_{l}\left(l\right)$.

We can finally obtain $P\left(\Omega,l\right)$ by Laplace inverting
$\hat{P}\left(\Omega,\lambda\right)$ in the appropriate scaling limit,
i.e. $\lambda\rightarrow0$, $\Gamma\rightarrow\infty$, and $a\lambda^{1/2}\Gamma\sim{\cal O}\!\left(1\right)$.
Thus, Eqs.~(\ref{eq:alfa}) and (\ref{eq:beta}) become\begin{eqnarray}
\alpha & = & a\sqrt{\lambda}\textrm{cosech}\left(a\sqrt{\lambda}\Gamma\right)~,\\
\beta & = & a\sqrt{\lambda}\coth\left(a\sqrt{\lambda}\Gamma\right)~,\end{eqnarray}
 respectively, yielding\begin{equation}
P\left(\Omega,l\right)=\frac{4\pi^{2}}{\Omega a^{2}\Gamma^{3}}\sum_{n=1}^{\infty}\left(-1\right)^{n+1}n^{2}\exp\left\{ -\left(\frac{n\pi}{a\Gamma}\right)^{2}l\right\} ~.\label{eq:P_omega_l2}\end{equation}

\section{Renormalization of a three-spin cluster\label{sec:Renormalization-of-a}}

A cluster of three spins $\mathbf{S}_{1}$, $\mathbf{S}_{2}$, and
$\mathbf{S}_{3}$, connected by antiferromagnetic XXZ couplings, can
be replaced at low energies by an effective spin-1/2 object. If the
three-spin Hamiltonian is given by\begin{eqnarray}
H_{123} & = & J\left(S_{1}^{x}S_{2}^{x}+S_{1}^{y}S_{2}^{y}+\Delta S_{1}^{z}S_{2}^{z}\right)\nonumber \\
 &  & +J\left(S_{2}^{x}S_{3}^{x}+S_{2}^{y}S_{3}^{y}+\Delta S_{2}^{z}S_{3}^{z}\right)~,\end{eqnarray}
 with $J>0$ and $0\leq\Delta\leq1$, then the ground state is doubly
degenerate. Thus, we can define an effective spin $\mathbf{S}_{0}$
such that, in the doublet subspace, the original spins are represented
by {}``weights'' defined by~\cite{vieira-aperiodic-PRB}\begin{equation}
S_{1}^{\alpha}=c_{1}^{\alpha}S_{0}^{\alpha},\quad S_{2}^{\alpha}=c_{2}^{\alpha}S_{0}^{\alpha},\quad S_{3}^{\alpha}=c_{3}^{\alpha}S_{0}^{\alpha}~,\end{equation}
 with $\alpha=x,\,\, y,\,\, z$ and \begin{equation}
c_{{\rm end}}^{x,y}\equiv c_{1}^{x,y}=c_{3}^{x,y}=\frac{\Delta+\sqrt{\Delta^{2}+8}}{2+\frac{1}{4}\left(\Delta+\sqrt{\Delta^{2}+8}\right)^{2}}~,\end{equation}
\begin{equation}
c_{\mathrm{end}}^{z}\equiv c_{1}^{z}=c_{3}^{z}=\frac{\frac{1}{4}\left(\Delta+\sqrt{\Delta^{2}+8}\right)^{2}}{2+\frac{1}{4}\left(\Delta+\sqrt{\Delta^{2}+8}\right)^{2}}~,\end{equation}
\begin{equation}
c_{\mathrm{mid}}^{x,y}\equiv c_{2}^{x,y}=-\frac{1}{1+\frac{1}{8}\left(\Delta+\sqrt{\Delta^{2}+8}\right)^{2}}~,\end{equation}
\begin{equation}
c_{\mathrm{mid}}^{z}\equiv c_{2}^{z}=-\frac{\frac{1}{4}\Delta\left(\Delta+\sqrt{\Delta^{2}+8}\right)}{1+\frac{1}{8}\left(\Delta+\sqrt{\Delta^{2}+8}\right)^{2}}~.\end{equation}

\section{Another derivation of the entanglement entropy in the strong-disorder
renormalization-group framework\label{sec:Another-derivation-of}}

Following Refael and Moore,~\cite{refael-moore} we \emph{exactly}
calculate the mean number of times a given bond is decimated, which
corresponds to the mean number of singlet links $M_{{\rm s}}$ crossing
a given point in the chain at the energy scale $\Omega$. Averaging
over all the sites in the lattice, this is simply the sum of the lengths
of all bonds decimated until the energy scale $\Omega$, divided by
the chain length:\begin{equation}
M_{{\rm s}}\left(\Omega\right)=\int_{\Omega}^{\Omega_{0}}{\rm d}\Omega\int_{0}^{\infty}{\rm d}l\, n_{\Omega}P\left(\Omega,l\right)l~.\label{eq:M-links}\end{equation}
 Using the results in Eqs.~(\ref{eq:n-omega-exact}) and (\ref{eq:P-omega-l})
we find that \begin{eqnarray}
M_{{\rm s}}\left(\Omega\right) & = & \frac{1}{3}\left[\ln\left(1+\vartheta_{0}\Gamma\right)+\frac{1}{1+\vartheta_{0}\Gamma}-1\right]\label{eq:Ms-1}\\
 & = & \frac{1}{3}\left(\frac{1}{2}\ln\frac{l}{l_{0}}+\sqrt{\frac{l_{0}}{l}}-1\right)~,\label{eq:Ms-2}\end{eqnarray}
 where we have used the relation between length and energy scales
in Eq.~(\ref{eq:l-medio-gama}). Again, the reader should be aware
of an extra 1/2 prefactor when integrating over $l$, due to the fact
that singlet lengths are restricted to odd multiples of $l_{0}$.
Finally, because any subsystem has two boundaries,\begin{equation}
S\left(l\right)=2s_{0}M_{s}=\frac{\gamma}{3}\ln\, l+b~,\end{equation}
 in which $\gamma=s_{0}=\ln\,2$ is a universal constant. Note that
the constant $b=-1/3+{\cal O}\!\left(l^{-1/2}\right)$ presented here
has no physical meaning. Although its value does not depend on the
initial disorder strength in this derivation, deviations from such
value are expected due to the spin clusters crossing the boundaries
(see Sec.~\ref{sec:Entanglement-entropy}).


\begin{thebibliography}{17}
\bibitem{foot-mccoy}There is an extra factor of 4 in Eq.~(6.9) of Ref.~\onlinecite{mccoy-68}.
\bibitem{foot-igloi}Note that in this case the recursive SDRG procedure is equivalent
to that of the critical point of the random Ising chain in a transverse
field, for which the coupling-constant distribution flow was exactly
obtained before by Igl\'{o}i and coworkers.~\cite{igloi-det-z-PRL,igloi-det-z-PRB}
\bibitem{foot-entropy}Notice we are using the logarithm in the natural basis in Eq.~(\ref{eq:S-definition}).
\end{thebibliography}
\end{document}